\documentstyle[12pt]{article}

\def\ar{\rightarrow}
\def\bib{\bibitem}

\def\intx{\int\! d^{\sl 4}x}
\def\intX{\int\! d^{\sl 4}X\,}

\def\intP{\int\! \frac{d^{\sl 4}P}{(2{\pi})^4}\,}

\def\lar{\longrightarrow}

\def\pa{\partial}
\def\rvec{\!\!\!\!^{^\rightarrow}}

\def\Tr{\,\mbox{Tr}\,}

\def\al{\alpha}
\def\be{\beta}
\def\ga{\gamma}
\def\de{\delta}
\def\ep{\varepsilon}

\def\la{\lambda}

\def\si{\sigma}
\def\om{\omega}

\def\Ga{{\it\Gamma}}
\def\La{{\it\Lambda}}
\def\Om{{\it\Omega}}

\def\beq{\begin{equation}}
\def\eeq{\end{equation}}
\def\bed{\begin{displaymath}}
\def\eed{\end{displaymath}}
\def\beqq{\begin{eqnarray}}
\def\eeqq{\end{eqnarray}}
\def\bedd{\begin{eqnarray*}}
\def\eedd{\end{eqnarray*}}

\begin{document}

\centerline{\normalsize\bf A RENORMALIZABLE THEORY OF QUANTUM GRAVITY:} \centerline{\normalsize\bf RENORMALIZATION PROOF OF THE GAUGE THEORY} \centerline{\normalsize\bf OF VOLUME PRESERVING DIFFEOMORPHISMS}

\vspace*{0.9cm}
\centerline{\footnotesize C. WIESENDANGER}
\baselineskip=12pt
\centerline{\footnotesize\it Aurorastr. 24, CH-8032 Zurich}
\centerline{\footnotesize E-mail: christian.wiesendanger@ubs.com}

\vspace*{0.9cm}
\baselineskip=13pt
\abstract{Inertial and gravitational mass or energy-momentum need not be the same for virtual quantum states. Separating their roles naturally leads to the gauge theory of volume-preserving diffeomorphisms of an inner four-dimensional space. The gauge-fixed action and the path integral measure occurring in the generating functional for the quantum Green functions of the theory are shown to obey a BRST-type symmetry. The related Zinn-Justin-type equation restricting the corresponding quantum effective action is established. This equation limits the infinite parts of the quantum effective action to have the same form as the gauge-fixed Lagrangian of the theory proving its spacetime renormalizability. The inner space integrals occurring in the quantum effective action which are divergent due to the gauge group's infinite volume are shown to be regularizable in a way consistent with the symmetries of the theory demonstrating as a byproduct that viable quantum gauge field theories are not limited to finite-dimensional compact gauge groups as is commonly assumed.}

\normalsize\baselineskip=15pt

\section{Introduction}
As of today a viable quantum field theory of gravitation has proven to be elusive \cite{car,clk}.

At the microscopic level the Standard Model (SM) of particle physics successfully describes the electromagnetic, weak and strong interactions based on quantized gauge field theories with the finite-dimensional compact gauge groups $U(1)$, $SU(2)$ and $SU(3)$ respectively \cite{stw1,stw2,cli,lor,stp,tpc}.

At the macroscopic level General Relativity (GR) successfully describes the gravitational interaction based on a classical gauge field theory with the non-compact diffeomorphism group of four-dimensional spacetime ${\bf R^{\sl 4}}$ as the gauge group \cite{stw3,lali,cmw}.

When trying to naively generalize the successful aspects of the gauge field theory ansatz to describe gravity at the quantum level one encounters unsurmountable difficulties - quantizing GR leads to a non-renormalizable theory and working with other finite-dimensional compact gauge groups to generalize the SM yields no description of gravity.

Whatever one tries one seems to bang one's head against two unremovable, yet intertwined roadblocks: (1) against GR and - underpinning it - the Principle of Equivalence stating that inertial and gravitational masses are equal which forces a geometric description of gravity and (2) against the seeming non-viability of gauge theories with non-compact gauge groups.

In this and a series of related papers \cite{chw1,chw2,chw3} we have systematically analyzed both roadblocks and (1) asked ourselves what physics at the quantum level we get when discarding the equality of inertial and gravitational mass for virtual quantum states and (2) made technical progress in formulating renormalizable gauge field theories based on a non-compact gauge group.

Before turning to the technical renormalization analysis let us illuminate both aspects a bit more in detail.

\subsection{Physical Aspect: Why should Inertial and Gravi- tational Mass be the Same for Virtual Quantum States?}

GR has been developed starting from the observed equality of inertial and gravitational mass $ m_I = m_G $. To be in agreement with observation this equality has to hold in any expression describing observable states in a gravitational context in their rest frames. However, in formulating a theory nothing enforces this equality for virtual (=non-observable) quantum states as long as it continues to hold for the on-shell (=observable) quantum states in that theory.

Now (a) the observed equality of inertial and gravitational mass of an on-shell physical object in its rest frame together with (b) the conservation of the inertial energy-momentum $ p_I^\mu $ in any frame tells us that in the rest frame
\beq \label{1}
p_I^\mu = (m_I,\underline{0}) =^{\!\!\!\!\!\!\! ^{ (a)}} (m_G,\underline{0}) = p_G^\mu
\eeq
assuming that the gravitational energy-momentum $ p_G^\mu $ plays a physical role different from that of the inertial energy-momentum, yet being observationally identical for on-shell objects. However, for off-shell states why shouldn't there be two separate conservation laws, one for the inertial energy-momentum and the other for the gravitational energy-momentum?

To explore this route let us postulate both $p_I^\mu$ and $p_G^\mu$ to be two separate four-vectors which are conserved, but in our approach through two different mechanisms. The conservation of $p_I^\mu$ is related to translation invariance in spacetime. Making use of Noether's theorem a second conserved four-vector can be constructed which is related to the invariance under volume-preserving diffeomorphisms of a four-dimensional inner space. That four-vector is then interpreted as the gravitational energy-momentum $p_G^\mu$ in the construction of a gauge theory of gravitation \cite{chw3}.

The observed equality of inertial and gravitational energy-momentum in this approach is assured by taking a gravitational limit for on-shell observable physical objects equating gravitational and inertial energy-momentum, the construction of which is based on the analysis of asymptotic states and the definition of a suitable $S$-matrix in the theory \cite{chw3}.

\subsection{Technical Aspect: Why should Viable Quantum Gauge Field Theories be Limited to Finite-Dimen-sional Compact Gauge Groups?}

In the process of constructing the gauge theory of volume-preserving diffeomorphisms of a four-dimensional inner space which emerges from the above thinking we have to deal with new difficulties arising from the non-compactness of the gauge group. (A) the gauge field Hamiltonian is not manifestly positive definite which is cured by a natural condition on the support of the gauge fields in inner space \cite{chw1}. (B) the quantization and subsequent derivation of Feynman rules yields no additional complications in comparison to the Yang-Mills case w.r.t. spacetime-related expressions, but it yields badly divergent-looking integrals over inner momenta related to the infinite volume of the gauge group - a phenomenon which does not plague Yang-Mills theories due to the assumed gauge group compactness. (C) as already mentioned a gravitational limit for on-shell observable physical objects has to be taken to ensure the observed equality of inertial and gravitational energy-momentum. This limit has to respect the unitarity of the physical $S$-matrix which we have established in \cite{chw3}.

In \cite{chw2} we have dealt with issues (A) and (B) at the one-loop level and established that the pure gauge field theory is asymptotically free whereas the inclusion of all SM fields destroys asymptotic freedom, hence assuring the observability of the gauge field quanta. To demonstrate the fundamental viability of the theory beyond one loop, however, we have to give both a proof of its renormalizability w.r.t the occurring spacetime divergences as well as to propose a general approach dealing with the inner divergences both of which we provide in this paper.

In the process we establish a well-defined perturbative expansion of the quantum effective action demonstrating its spacetime renormalizability and the existence of regularization schemes for the inner momentum integrals consistent with the symmetries of the theory. In fact, the spacetime-renormalized and inner-space-regularized quantum effective action {\it defines} a viable quantum field theory for each regularization scheme in terms of the original finite number of coupling constants, masses etc.

Stated otherwise it is possible to consistently establish a quantum gauge field theory not only for compact gauge groups, but also for at least the non-compact gauge group of volume-preserving diffeomorphisms. The price to pay comes in the form of an additional regularization scheme for the divergent sums over inner degrees of freedom - each such scheme which is compatible with the symmetries of the theory establishes one well-defined version of the quantum theory belonging to the classical gauge theory one starts with. And each such version with its finite number of coupling constants, masses etc. yields as precise predictions as do its Yang-Mills cousins - predictions which are equal for all such versions at tree level, but depend on the chosen regularization scheme for the loop contributions. In exactly the same way as experiment has to tell the physical values of the various couplings, masses etc. in this or the Yang-Mills cases, experiment ultimately has then to choose the regularization preferred by Gravity.

So let us turn to implement the program outlined above.

\section{Classical Gauge Theory of Volume-Preser- ving Diffeomorphisms}

In this section we review the basics of the gauge theory of volume preserving diffeomorphisms as presented in \cite{chw1}.

Let us start with a four-dimensional real vector space
${\bf R}^{\sl 4}$ with elements labelled $X^\al$ without a metric structure at this point which we will call inner space in the following. Volume-preserving diffeomorphisms 
\beq \label{2}
X^\al\lar X'^\be = X'^\be(X^\al),\:\:\al,\be=\sl{0,1,2,3}
\eeq
act as a group ${\overline{DIFF}}\,{\bf R}^{\sl 4}$ under composition on this space. $X'^\be (X)$ denotes an invertible and differentiable coordinate transformation of ${\bf R\/}^{\sl{4}}$ with unimodular Jacobian
\beq \label{3}
\det\left(\frac{\pa X'^\be (X)}{\pa X^\al}\right) = 1.
\eeq

Next we want to represent this group on spaces of differentiable functions $\psi(X)$ which will in differential-geometrical terms later serve as fibres and sections respectively of the vector bundles to be constructed. To be specific we require the $\psi(X)$ to be differentiable functions on the inner space introduced above and to be integrable such that the scalar product
\beq \label{4}
\langle \psi \!\mid\! \chi \rangle \equiv \intX \La^{-4} \psi^\dagger (X) \cdot\chi(X)
\eeq
is well-defined. Above we have introduced a parameter $\La$ of dimension length, $[\La]=[X]$, so as to define a dimensionless scalar product.

Turning to the passive representation of ${\overline{DIFF}}\,{\bf R}^{\sl 4}$ in field space for infinitesimal transformations $X'^\al (X) = X^\al + \La\, {\cal E}^\al (X)$ with ${\cal E}^\al$ dimensionless we have
\beqq \label{5}
X^\al &\lar& X'^\al = X^\al, \\
\psi (X) &\lar& \psi'(X) = \psi(X) -\, {\cal E}^\al (X)\cdot \La\, \nabla_\al\,\psi(X) \nonumber
\eeqq
transforming the fields only, where $\nabla_\al = \frac{\pa\,\,\,}{\pa X^\al}$ represents partial differentiation in inner space.

The unimodularity condition Eqn.(\ref{3}) translates into the infinitesimal gauge parameter ${\cal E}^\al$ being divergence-free
\beq \label{6}
\nabla_\al {\cal E}^\al (X) = 0. \nonumber
\eeq
Note the crucial fact that the algebra ${\overline{\bf diff}}\,{\bf R}^{\sl 4}$ of the divergence-free ${\cal E}$s closes under commutation. For $\nabla_\al {\cal E}^\al (X) = \nabla_\be {\cal F}^\be (X) = 0$ we have
\beq \label{7} 
\left[{\cal E}^\al (X) \cdot \nabla_\al, {\cal F}^\be (X) \cdot \nabla_\be \right] = \left( {\cal E}^\al (X) \cdot \nabla_\al {\cal F}^\be (X)
- {\cal F}^\al (X) \cdot \nabla_\al {\cal E}^\be (X) \right) \nabla_\be
\eeq
with
\beq \label{8} 
\nabla_\be \left({\cal E}^\al (X) \cdot \nabla_\al {\cal F}^\be (X) 
- {\cal F}^\al (X) \cdot \nabla_\al {\cal E}^\be (X) \right) = 0
\eeq
as required by the finite transformations ${\overline{DIFF}}\,{\bf R}^{\sl 4}$ forming a group under composition.

As a result we can write infinitesimal transformations in field space 
\beq \label{9}
U_{\cal E} (X) \equiv {\bf 1} - {\cal E}(X),\:\: {\cal E}(X)= {\cal E}^\al (X)\cdot \La\, \nabla_\al
\eeq
as anti-unitary operators w.r.t. the scalar product Eqn.(\ref{4}). Both the ${\cal E}(X)$ and the $\nabla_\al$ are anti-hermitean w.r.t. the scalar product Eqn.(\ref{4}). The decomposability of ${\cal E}(X)$ w.r.t. to the operators $\nabla_\al$ will be crucial for the further development of the theory, especially for identifying the gauge field variables of the theory. 

Introducing the variation $\de_{_{\cal E}} ..\equiv ..' - ..$ of an expression under a gauge transformation we can finally write
\beq \label{10}
\de_{_{\cal E}}\psi (X) \equiv \psi' (X) - \psi (X) = -\, {\cal E}^\al (X) \cdot \La\, \nabla_\al\,\psi (X). \eeq

In mathematical terms we have just reviewed the group ${\overline{DIFF}}\,{\bf R}^{\sl 4}$, the algebra ${\overline{\bf diff}}\,{\bf R}^{\sl 4}$ and defined their representations on a suitable space of functions. Note that spacetime has played no role so far.

Next we turn to the four-dimensional Minkowski spacetime ({\bf M}$^{\sl 4}$,\,$\eta$) with metric $\eta=\mbox{diag}(-1,1,1,1)$ and elements labelled $x^\mu$ which will serve as the base space of the vector bundles we construct.

Extending the global volume-preserving diffeomorphism group to a group of local transformations we allow ${\cal E}^\al (X)$ to vary with $x$ as well, i.e. we allow for $x$-dependent volume-preserving infinitesimal gauge parameters ${\cal E}^\al (X)\ar {\cal E}^\al (x,X)$. To represent the group we have to allow the functions $\psi(X)$ to become $x$-dependent fields $\psi(x,X)$ as well.

We note that these fields $\psi(x,X)$ might live in non-trivial representation spaces of both the spacetime Lorentz group with spin $s\not= 0$ and of other symmetry groups such as $SU(N)$. These representations factorize w.r.t the inner diffeomorphism group representations we introduce below which is consistent with the Coleman-Mandula theorem.

In generalization of Eqn.(\ref{9}) we thus consider
\beq \label{11}
U_{\cal E} (x,X) \equiv {\bf 1} - {\cal E}(x,X),\:\: {\cal E}(x,X)= {\cal E}^\al (x,X)\cdot \La\, \nabla_\al.
\eeq
The formulae Eqns.(\ref{5}) together with Eqn.(\ref{6}) with the fields now $x$-depen- dent as well still define the representation of the volume-preserving diffeomorphism group in field space.

Next we introduce a covariant derivative $D_\mu$ which is defined by the transformation requirement
\beq \label{12}
D'_\mu (x,X) \,\, U_{\cal E} (x,X) = U_{\cal E} (x,X)\,\, D_\mu (x,X),
\eeq
where $D'_\mu (x,X)$ denotes the gauge-transformed covariant derivative.

To fulfil Eqn.(\ref{12}) we make the usual ansatz
\beq \label{13}
D_\mu (x,X) \equiv \pa_\mu + A_\mu (x,X),\quad A_\mu (x,X) \equiv A_\mu\,^\al (x,X)\cdot \La\, \nabla_\al
\eeq
decomposing $A_\mu (x,X)$ w.r.t the generators $\nabla_\al$ of the diffeomorphism algebra in field space. In order to have the gauge fields in the algebra ${\overline{\bf diff}}\,{\bf R}^{\sl 4}$ we impose in addition
\beq \label{14}
\nabla_\al A_\mu\,^\al (x,X) = 0
\eeq
consistent with $\nabla_\be {\cal E}^\be (x,X)=0$. As a consequence the usual ordering problem for $A_\mu\,^\al$ and $\nabla_\al$ in the definition of $D_\mu$ does not arise and $D_\mu$ is anti-hermitean w.r.t to the scalar product
\beq \label{15}
\langle \psi \!\mid\! \chi \rangle \equiv \intx\intX \La^{-4} \psi^\dagger (x,X) \cdot\chi(x,X).
\eeq 

The requirement Eqn.(\ref{12}) translates into the transformation law for the gauge field
\beq \label{16}
\de_{_{\cal E}} A_\mu (x,X) = \pa_\mu {\cal E} (x,X) -  \left[{\cal E} (x,X), A_\mu (x,X) \right] \eeq
which reads in components
\beqq \label{17}
\de_{_{\cal E}} A_\mu\,^\al (x,X) &=& \pa_\mu {\cal E}^\al (x,X) 
+ A_\mu\,^\be (x,X) \cdot \La\, \nabla_\be {\cal E}^\al (x,X) \nonumber \\
&-& {\cal E}^\be (x,X) \cdot \La\, \nabla_\be A_\mu\,^\al (x,X)
\eeqq
respecting $\nabla_\al \de_{_{\cal E}} A_\mu\,^\al = 0$.

Note that the consistent decomposition of both $A_\mu$ and $A'_\mu$ w.r.t. the generators $\nabla_\al$ is crucial for the theory's viability. It is ensured by the closure of the algebra Eqn.(\ref{7}) and the gauge invariance of $\nabla_\al A_\mu\,^\al =0$ for gauge parameters fulfilling $\nabla_\be {\cal E}^\be =0$ also when $x$-dependent.

Let us next define the field strength operator $F_{\mu\nu}$ in the usual way
\beqq \label{18}
F_{\mu\nu}(x,X)&\equiv& \left[D_\mu (x,X), D_\nu (x,X) \right] \nonumber \\
&=& F_{\mu\nu}\,^\al (x,X)\cdot \La\, \nabla_\al 
\eeqq
which again can be decomposed consistently w.r.t. $\nabla_\al$.
The field strength components $ F_{\mu\nu}\,^\al (x,X)$ are calculated to be
\beqq \label{19}
F_{\mu\nu}\,^\al (x,X)
&\equiv& \pa_\mu A_\nu\,^\al (x,X) - \pa_\nu A_\mu\,^\al (x,X) \nonumber \\
&+& A_\mu\,^\be (x,X)\cdot \La\, \nabla_\be A_\nu\,^\al (x,X) \\
&-& A_\nu\,^\be (x,X)\cdot \La\, \nabla_\be A_\mu\,^\al (x,X). \nonumber
\eeqq
Under a local gauge transformation the field strength and its components transform covariantly
\beqq \label{20}
\de_{_{\cal E}} F_{\mu\nu} (x,X)
&=& - \left[{\cal E} (x,X) , F_{\mu\nu} (x,X) \right], \nonumber \\
\de_{_{\cal E}} F_{\mu\nu}\,^\al (x,X) 
&=& F_{\mu\nu}\,^\be (x,X) \cdot \La\, \nabla_\be {\cal E}^\al (x,X) \\
&-& {\cal E}^\be (x,X) \cdot \La\, \nabla_\be F_{\mu\nu}\,^\al (x,X). \nonumber
\eeqq
As required for algebra elements $\nabla_\al F_{\mu\nu}\,^\al = 0$ and $\nabla_\al \de_{_{\cal E}} F_{\mu\nu}\,^\al = 0$ for gauge fields fulfilling $\nabla_\al A_\mu\,^\al = 0$ and gauge parameters fulfilling $\nabla_\be {\cal E}^\be =0$.

The above can be viewed as giving rise to a principal bundle {\bf P} with base space {\bf M}$^{\sl 4}$, the typical fibre given by the volume-preserving diffeomorphisms $X'^\be (X)$ of ${\bf R}^{\sl 4}$ which is identical to the structure group given by ${\overline{DIFF}}\,{\bf R}^{\sl 4}$. The gauge field ${\cal A} = A_\mu\, dx^\mu$ is then the local form of the connection and the field strength ${\cal F} = d{\cal A} + {\cal A} \wedge {\cal A} = \frac{1}{2} F_{\mu\nu}\, dx^\mu \wedge dx^\nu $ the local form of the curvature of the principal bundle {\bf P} ({\bf M}$^{\sl 4}$, ${\overline{DIFF}}\,{\bf R}^{\sl 4}$). Locally the mathematical existence of the gauge field and field strength is hence assured. Finally in this paper we do not address questions about the global structure of such a principal bundle.

Note that no reference to a metric in inner space has been necessary so far.

Turning to the dynamics of the classical gauge theory the Lagrangian for the gauge fields as given in \cite{chw1} does depend on a metric $g_{\al\be}(x,X)$ of Minkowskian signature on the inner space ${\bf R}^{\sl 4}$ which has been introduced in \cite{chw1} and discussed there. Under inner coordinate transformations this metric transforms as a contravariant tensor
\beqq \label{21}
\de_{_{\cal E}} g_{\al\be} (x,X)
&=& - {\cal E}^\ga (x,X)\cdot \La\, \nabla_\ga g_{\al\be} (x,X)
\nonumber \\
&-& g_{\ga\be} (x,X)\cdot \La\, \nabla_\al {\cal E}^\ga (x,X) \\
&-& g_{\al\ga} (x,X)\cdot \La\, \nabla_\be {\cal E}^\ga (x,X). \nonumber
\eeqq
In general inner coordinates the Lagrangian is given by \cite{chw1}
\beq \label{22}
{\cal L} (x,X) = -\frac{1}{4}\, F_{\mu\nu}\,^\al(x,X)\cdot F^{\mu\nu}\,_\al (x,X)
\eeq
and is indeed covariant under the combined gauge transformations Eqn.(\ref{20}) and Eqn.(\ref{21}). Above, inner indices such as $\al$ are raised and lowered with $g_{\al\be} (x,X)$.

Excluding any dynamical role for the metric - which we take as an a priori as explained in \cite{chw1} - we require that the geometry of the inner space is flat, hence Riem$(g) = 0$. This means that it is always possible to choose global Cartesian coordinates with the metric $g_{\al\be}(x,X) = \eta_{\al\be}$ collapsing to the global Minkowski metric $\eta$ in inner space.

Such choices of coordinates amount to partially fixing a gauge the class of which are the Minkowski gauges \cite{chw1}. The remaining gauge degrees of freedom leaving $\eta$ invariant restrict the infinitesimal gauge parameters to the form
\beq \label{23}
{\cal E}^\al (x,X) = \ep^\al (x)  + \om^\al\,_\be (x) X^\be
\eeq 
with $\ep (x)$ and $\om (x)$ constant in inner space and $\om_{\al\be} (x)$ antisymmetric in $\al$, $\be$ which is the (infinitesimal) Poincar\'e group acting on inner space.

In Minkowski gauges and Cartesian inner coordinates the Lagrangian is \cite{chw1} still given by Eqn.(\ref{22}) with inner indices such as $\al$ raised and lowered with $ \eta_{\al\be}$. In such gauges, however, it is only covariant under Poincar\'e transformations as given in Eqn.(\ref{23}) above.

\section{Generating Functional for the Green Functions
of the Gauge Theory of Volume-Preser- ving Diffeomorphisms}

In this section we establish the starting point for the renormalizability proof of the gauge theory of volume preserving diffeomorphisms in terms of the gauge-fixed generating functional for its quantum Green functions as presented in detail in \cite{chw2}.

After the review of the classical theory let us turn to the quantum theory and specifically to the generating functional ${\cal Z}\left[\eta, J \right]$ for the quantum Green functions of the gauge theory of volume-preserving diffeomorphisms given in terms of the gauge fields  $A_\mu\,^\al (x,X)$, the ghost and anti-ghost fields $\om^*_\be (x,X)$, $\om^\de (x,X)$ and a matter field $\psi (x,X)$ as established in detail in \cite{chw2} 
\beqq \label{24}
{\cal Z}\left[\eta, J \right]
&=& \int\Pi_{\!\!\!\!\!\!_{_{_{x,X}}}} d\psi \cdot
\int\Pi_{\!\!\!\!\!\!_{_{_{x,X;\mu,\al}}}}
\!\!\!\!\!\!\!\!dA_\mu\,^\al \;
\Pi_{\!\!\!\!\!_{_{_{\mu}}}} \,\,
\de (\nabla_\al A_\mu\,^\al) \nonumber \\
& & \!\!\!\!\!\!\!\!\!\!\!\!\!\!\!\!\!\!
\cdot \int\,\Pi_{\!\!\!\!\!\!_{_{_{x,X;\ga}}}}\!\!d\om^*_\ga \;
\de (\nabla^\ga \om^*_\ga)
\cdot \int\,\Pi_{\!\!\!\!\!\!_{_{_{x,X;\de}}}}\!\!d\om^\de \;
\de (\nabla_\de\, \om^\de) \\
& & \!\!\!\!\!\!\!\!\!\!\!\!\!\!\!\!\!\!
\cdot \exp\,i\,\left\{S_{MOD} + S_M + \int J\cdot A + \int \eta\cdot \psi + \ep \mbox{-terms} \right\}. \nonumber
\eeqq
The path integrals over the fields are restricted by the $\de$-functions to fields living in the gauge algebra. For the integrals over spacetime and inner space we have chosen the shorthand notation
\beq \label{25}
\int\dots \sim \intx \intX \La^{-4}\dots
\eeq
which we will keep throughout the rest of the paper and we have dropped the explicit dependence on the coordinates $x,X$ which we will do from now on throughout this paper.

The gauge-fixed action $S_{MOD}$ related to the gauge and ghost fields is written in terms of a Lagrangian density ${\cal L}_{MOD}$
\beqq \label{26}
S_{MOD} &=& \int\, {\cal L}_{MOD} \\
{\cal L}_{MOD} &\equiv& {\cal L} + {\cal L}_{GF} + {\cal L}_{GH}, \nonumber
\eeqq
where the latter is a sum of the gauge field Lagrangian ${\cal L}$ from Eqn.(\ref{22}) covariant under the combined gauge transformations Eqn.(\ref {20}) and Eqn.(\ref {21}), a gauge-fixing term ${\cal L}_{GF}$ and the ghost field Lagrangian ${\cal L}_{GH}$
\beqq \label{27}
{\cal L}_{GF} &\equiv& - \frac{1}{2\xi} \, f_\al [A] \cdot f^\al [A] \\
{\cal L}_{GH} &\equiv& \om^*_\al \cdot
{\cal F}^\al\,_\be \left[A \right] \om^\be. \nonumber
\eeqq
In general, the inner indices in Eqns.(\ref{27}) are contracted with $g$.

To represent matter we have added a Dirac field with action $S_M$
\beqq \label{28}
S_M &=& \int\, {\cal L}_M \\
{\cal L}_M  (x,X)&=&
- \bar\psi\, \ga^\mu (\pa_\mu\rvec + A_\mu\,^\al \cdot \La\, {\nabla\rvec}_\al) \psi - \, m\, \bar\psi\, \psi. \nonumber
\eeqq
Finally we have introduced currents $J$ and $\eta$ coupled to the gauge and matter fields.

Above ${\cal F}^\al\,_\be \left[A \right]$ denotes the Faddeev-Popov-DeWitt kernel
\beq \label{29}
{\cal F}^\al\,_\be \left[A \right] \equiv
\frac{\de f^\al [A_{_{\cal E}}]}{\de\, {\cal E}^\be} _{\mid_{_{{\cal E}=0}}}
\eeq 
belonging to the gauge-fixing functional $f^\al [A]$ with $A_{_{\cal E}} \equiv A + \de_{_{\cal E}} A$.

Note that $S_{MOD}$ is not invariant under the joint gauge transformations Eqns.(\ref{10}) and (\ref{17}) and Eqn.(\ref{21}). As we will see below, however, $S_{MOD}$ and the quantum theory feature another invariance which is a remnant of the gauge invariance of the classical theory.

To be specific in our further analysis let us (i) partially fix the gauge to Minkowski gauges introduced above which preserve the inner metric $\eta$ (see Appendix and \cite{chw1}) and (ii) fix the remaining gauge degrees of freedom by choosing the Lorentz gauge taking $f^\al [A]$ to be
\beq \label{30}
f^\al [A] \equiv \pa^\mu A_\mu\,^\al.
\eeq
The Faddeev-Popov-DeWitt kernel for this choice is easily calculated to be
\beq \label{31}
{\cal F}^\al\,_\be \left[A\right] = \pa^\mu\! \left(\pa_\mu \,\eta^\al\,_\be + A_\mu\,^\ga\cdot \La\, \nabla_\ga \,\eta^\al\,_\be - \La\, \nabla_\be A_\mu\,^\al\right) 
\eeq
and ${\cal L}_{MOD}$ becomes
\beqq \label{32}
{\cal L}_{MOD} &=& - \frac{1}{4}\, F_{\mu\nu}\,^\al \cdot F^{\mu\nu}\,_\al, \nonumber \\
&-& \frac{1}{2\xi} \, \pa_\mu A^\mu\,_\al \cdot \pa^\nu A_\nu\,^\al \\
&-& \pa_\mu \om^*_\al \cdot \pa^\mu \om^\al \nonumber \\
&-& \pa_\mu \om^*_\al \cdot
\left(A^\mu\,_\be\cdot \La\, \nabla^\be \om^\al - \om^\be\cdot \La\, \nabla_\be A^\mu\,^\al\right). \nonumber
\eeqq
Due to (ii) the inner indices in Eqn.(\ref{32}) are contracted with $\eta$ as is the case throughout the rest of the paper.

Denoting the mass dimension of a field determined as usual by the quadratic part of the Lagrangian by $[..]$ we find 
\beqq \label{33}
& & [ A_\mu\,^\al ] = [ \om^\al ] = [ \om^*_\al ] = 1,
\nonumber \\
& & [ \psi ] = 3/2.
\eeqq
Hence, by inspection the Lagrangian above contains products of fields and their spacetime derivatives with mass dimensions four or less only. This ensures renormalizability in the Dyson sense, i.e. counterterms which have to be introduced to absorb spacetime divergences arising in perturbation theory have mass dimensions four or less as well. For the theory to be truly renormalizable, however, these counterterms arising in a perturbation expansion of the generating functional must be shown to take the same form as the above Lagrangian - a task to which we turn next.

Note that we will have to deal with an additional type of divergences in Feynman graphs arising from the generalized sums over inner degrees of freedom. These turn out to be divergent integrals over inner momentum space variables which we will properly define in section 10.

\section{BRST-Type Invariance of Modified Gauge-Fixed Action and of Path Integral Measure occurring in the Generating Functional for the Quantum Green Functions}

In this section we rewrite the generating functional in terms of Nakanishi-Lautrup fields and a new action $S_{NEW}$. We then demonstrated the invariance of both $S_{NEW}$ and the path integration measure under a BRST-type symmetry.

Let us start with Eqn.(\ref{26})
\beq \label{34}
S_{MOD} = S - \frac{1}{2\xi}\, \int\,f_\al \cdot f^\al
+ \int\, \om^*_\al \cdot {\it\Delta}^\al,
\eeq
where we have introduced the quantity
\beq \label{35}
{\it\Delta}^\al \equiv {\cal F}^\al \,_\be \cdot \om^\be.
\eeq 

Re-expressing
\beqq \label{36}
& & \exp\left\{ - \frac{i}{2\xi}\, \int\, f_\al \cdot f^\al \right\} 
\propto \int\Pi_{\!\!\!\!\!\!_{_{_{x,X;\al}}}}\!\!\!\!dh^\al \;
\de (\nabla_\al h^\al) \\
& &\quad\quad\quad \cdot \exp\left\{ \frac{i\, \xi}{2}\, \int\,h_\al \cdot h^\al + i\, \int\,h_\al \cdot f^\al \right\} \nonumber 
\eeqq
in terms of a Gaussian integral over the Nakanishi-Lautrup fields $h^\al$ and introducing the corresponding new modified action
\beq \label{37}
S_{NEW} = S + \int\, \om^*_\al \cdot {\it\Delta}^\al + \int\,h_\al \cdot f^\al + \frac{\xi}{2} \, \int\,h_\al \cdot h^\al
\eeq
the Green functions of the theory are now given as path integrals over the fields $A$, $\om^*$, $\om$, $h$, $\psi$ with weight $\exp\, i \, \{S_{NEW}+S_M\}$.

By construction the gauge-fixed modified action $S_{NEW}$ is not invariant under gauge transformations. However, it is invariant under BRST-type transformations parametrized by an infinitesimal fermionic $\theta$ anticommuting with ghost and fermionic fields. The BRST-type variations in the theory of volume-preserving diffeomorphisms are given by
\beqq \label{38}
\de_\theta A_\mu\,^\al &=& \theta\left(\pa_\mu \om^\al
+ A_\mu\,^\be\cdot \La\, \nabla_\be \om^\al -\om^\be\cdot \La\, \nabla_\be A_\mu\,^\al \right)  \nonumber \\
\de_\theta \om^*_\al &=& - \theta\, h_\al \nonumber \\
\de_\theta \om^\al &=& - \theta\, \om^\be\cdot \La\, \nabla_\be \om^\al \\
\de_\theta h_\al &=& 0 \nonumber \\
\de_\theta \psi &=& - \theta\, \om^\be\cdot \La\, \nabla_\be \psi. \nonumber
\eeqq
The transformations Eqns.(\ref{38}) are nilpotent, i.e. if ${\cal F}$ is any functional of $A$, $\om^*$, $\om$, $h$, $\psi$ and we define $s{\cal F}$ by
\beq \label{39}
\de_\theta {\cal F} \equiv \theta s{\cal F}
\eeq
then  
\beq \label{40}
\de_\theta s{\cal F} = 0 \quad\mbox{or}\quad s(s{\cal F}) = 0.
\eeq 
The proof for the fields above is straightforward, but somewhat tedious. Here we just sketch the verification of $s(s A_\mu\,^\al)=0$
\beqq \label{41}
\!\!\!\!\!\!\!\!
\de_\theta sA_\mu\,^\al &=& \theta\, \Bigg\{
\pa_\mu \left( - \om^\be\cdot \La\, \nabla_\be \om^\al \right) \nonumber \\
&+& \left(\pa_\mu \om^\be
+ A_\mu\,^\ga\cdot \La\, \nabla_\ga \om^\be -\om^\ga\cdot \La\, \nabla_\ga A_\mu\,^\be \right)\cdot \La\, \nabla_\be \om^\al \nonumber \\
&-& A_\mu\,^\be\cdot \La\, \nabla_\be \left( \om^\ga\cdot \La\, \nabla_\ga \om^\al \right)
+ \left( \om^\ga\cdot \La\, \nabla_\ga \om^\be \right)\cdot \La\, \nabla_\be A_\mu\,^\al \\
&+& \om^\be\cdot \La\, \nabla_\be \left( \pa_\mu \om^\al 
+ A_\mu\,^\ga\cdot \La\, \nabla_\ga \om^\al - \om^\ga\cdot \La\, \nabla_\ga A_\mu\,^\al \right)\Bigg\} \nonumber \\
&=& 0 \nonumber \eeqq
using the chain-rule and the anticommutativity of $\theta$ with $\om$. As a result we have
\beqq \label{42}
s(s A_\mu\,^\al) &=& 0,\quad s(s\, \om^*_\al) = 0, \quad s(s\, \om^\al) = 0 \nonumber \\
s(s\, h_\al) &=& 0, \quad s(s\, \psi) = 0.
\eeqq
The extension to products of polynomials in these fields follows easily \cite{stw2}.

To verify the BRST invariance of $S_{NEW}$ we note that the BRST transformation acts on functionals of matter and gauge fields as a gauge transformation with gauge parameter ${\cal E}_\al = \theta\, \om_\al$. Hence
\beq \label{43}
\de_\theta S = 0 \quad \mbox{and} \quad \de_\theta S_M = 0.
\eeq
Next with the use of Eqn.(\ref{29}) we determine the BRST transform of $f^\al$
\beq \label{44}
\de_\theta f^\al = \frac{\de f^\al}{\de\, {\cal E}_\be} _{\mid_{_{{\cal E}=0}}} \!\!\!\!\!\!\!\!\cdot \theta \, \om_\be= \theta \, {\it\Delta}^\al
\eeq
which yields
\beq \label{45}
-s\left(\om^*_\al \cdot f^\al + \frac{\xi}{2} \, \om^*_\al \cdot h^\al \right) = \om^*_\al \cdot {\it\Delta}^\al + h_\al \cdot f^\al + \frac{\xi}{2} \, h_\al \cdot h^\al.
\eeq

Hence we can rewrite
\beq \label{46}
S_{NEW} = S + s{\it\Psi},
\eeq
where
\beq \label{47}
{\it\Psi} \equiv - \int \left(\om^*_\al \cdot f^\al + \frac{\xi}{2} \, \om^*_\al \cdot h^\al \right).
\eeq
Finally it follows from the nilpotency of the BRST transformation
\beq \label{48}
\de_\theta S_{NEW} = 0.
\eeq

Next let us analyze the path integration measure
\beqq \label{49}
& & \Pi_{\!\!\!\!\!\!_{_{_{x,X}}}} d\psi \cdot
\Pi_{\!\!\!\!\!\!_{_{_{x,X;\mu,\al}}}}
\!\!\!\!\!\!\!\!dA_\mu\,^\al \;
\Pi_{\!\!\!\!\!_{_{_{\mu}}}} \,\,
\de (\nabla_\al A_\mu\,^\al)
\cdot \Pi_{\!\!\!\!\!\!_{_{_{x,X;\be}}}}\!\!d h^\be \;
\de (\nabla^\be\, h_\be) \\
& &\quad\quad
\cdot \Pi_{\!\!\!\!\!\!_{_{_{x,X;\ga}}}}\!\!d\om^*_\ga \;
\de (\nabla^\ga \om^*_\ga)
\cdot \Pi_{\!\!\!\!\!\!_{_{_{x,X;\de}}}}\!\!d\om^\de \;
\de (\nabla_\de\, \om^\de). \nonumber
\eeqq

Under the BRST-type transformations Eqns.(\ref{38}) the Jacobian turns out to be \cite{jez}
\beq \label{50}
{\cal J} = 1 + \Tr \left( 
\frac{\de (\de_\theta A_\mu\,^\al) }{\de A_\nu\,^\be } -
\frac{\de(\de_\theta \, \om^*_\al, \de_\theta \, \om^\be, \de_\theta \, \psi)}
{\de(\om^*_\be, \om^\de, \psi)}
\right) = 1
\eeq
as the trace is easily shown to vanish.

In addition the BRST-type transformations Eqns.(\ref{38}) respect the di- vergence-free condition ensuring that the fields live in the gauge algebra
\beqq \label{51}
\nabla_\al \de_\theta A_\mu\,^\al &=& 0 \nonumber \\
\nabla^\al \de_\theta \om^*_\al &=& 0 \\
\nabla_\al \de_\theta \om^\al &=& 0 \nonumber \\
\nabla^\al \de_\theta h_\al &=& 0. \nonumber
\eeqq

As a result the measure Eqn.(\ref{49}) is invariant under the BRST-type transformations Eqns.(\ref{38}).

\section{Symmetries of the Quantum Effective Action and the Zinn-Justin Equation}

In this section we derive the Zinn-Justin equation for the gauge theory of volume-preserving diffeomorphisms which follows from the BRST-type invariance of both the modified action $S_{NEW}$ and the path integral measure.

We start with the generating functional for Green functions ${\cal Z}\left[J, K \right]$ in the presence of additional currents $K_n$ coupled to the nilpotent BRST-type transformations $\Delta^n$ chosing the compact notation
\beqq \label{52}
{\cal Z}\left[J, K \right]
&\equiv& \int\Pi_{\!\!\!\!\!\!_{_{_{x,X;n}}}}
\!\!\!\!d\chi^n (x,X)\, \Pi_{\!\!\!\!\!_{_{_{n}}}}\,
D \left[g \left[x,X;\chi^n \right] \right] \\
& & \!\!\!\!\!\!\!\!\!\!\!\!\!\!\!\!\!\!
\cdot \exp\,i\,\left\{S_{NEW} + S_M + \int \Delta^n\cdot K_n + \int \chi^n\cdot J_n + \ep \mbox{-terms} \right\}, \nonumber
\eeqq
where
\beq \label{53}
\chi^n_{\,\al} \equiv A^{\sl 1}\,_{\!\al}, A^{\sl 2}\,_{\!\al}, 
A^{\sl 3}\,_{\!\al}, A^{\sl 4} \,_{\!\al};\; \om^*_\al;\; \om_\al;\; h_\al
\eeq
denote the gauge and ghost field variables constrained to live in the gauge algebra which is ensured by
\beqq \label{54}
g \left[x,X;\chi^n \right] &=& \nabla^\al\, \chi^n_{\,\al} (x,X) \\
D \left[g (x,X) \right] &=& \de (g(x,X))\nonumber
\eeqq
and where
\beq \label{55}
\chi^{\sl 8} \equiv \psi
\eeq 
denotes the matter field. The BRST-type transformations
\beq \label{56}
\de_\theta \chi^n (x,X) = \theta \Delta^n \left[x,X;\chi \right]
\eeq
act on the various fields as defined by Eqns.(\ref{38})
\beq \label{57}
\begin{array}{clcl}
\Delta^n &= \pa_\mu \om^\al + A_\mu\,^\be\cdot \La\, \nabla_\be \om^\al
- \om^\be\cdot \La\, \nabla_\be A_\mu\,^\al &\mbox{for}
& \quad\chi^n = A_\mu\,^\al \nonumber \\
\Delta^n &= - h_\al & \mbox{for} 
& \quad \chi^n = \om^*_\al \\
\Delta^n &= - \om^\be\cdot \La\, \nabla_\be \om^\al & \mbox{for} 
& \quad \chi^n = \om^\al \nonumber \\
\Delta^n &= 0 & \mbox{for} & \quad \chi^n = h^\al \nonumber \\
\Delta^n &= - \om^\be\cdot \La\, \nabla_\be \psi & \mbox{for} 
& \quad \chi^n = \psi. \nonumber
\end{array}
\eeq

As shown in the previous section all: the modified action $S_{NEW}$, the matter action $S_M$, the nilpotent BRST-transformations $\Delta^n$ and the path integration measure in ${\cal Z}\left[J, K \right]$ are invariant under the BRST-type transformations Eqns.(\ref{57}). Hence we obtain
\beqq \label{58}
{\cal Z}\left[J, K \right]
&=& \int\Pi_{\!\!\!\!\!\!_{_{_{x,X;n}}}}
\!\!\!\!d\left(\chi^n + \theta \Delta^n \left[\chi \right] \right)\, 
\Pi_{\!\!\!\!\!_{_{_{n}}}}\, D \left[g \left[\chi^n 
+ \theta \Delta^n \left[\chi \right] \right] \right] \nonumber \\
&\cdot& \!\!\!\!\! \exp\,i\,\bigg\{
S_{NEW} \left[\chi^n + \theta \Delta^n \left[\chi \right] \right]  + 
S_M \left[\chi^n + \theta \Delta^n \left[\chi \right] \right] \nonumber \\
& &  + \int \Delta^n \left[\chi^m + \theta \Delta^m 
\left[\chi \right] \right] \cdot K_n + 
\int \left(\chi^n + \theta \Delta^n \left[\chi \right] \right) \cdot J_n
\bigg\} \\
&=& {\cal Z}\left[J, K \right]
+\, i\, \theta \int\Pi_{\!\!\!\!\!\!_{_{_{x,X;n}}}}\!\!\!\! d\chi^n\, 
\Pi_{\!\!\!\!\!_{_{_{n}}}}\, D \left[g \left[\chi^n \right] \right]  \cdot 
\left( \int \Delta^m \left[\chi \right] \cdot J_m \right) \nonumber \\
&\cdot& \!\!\!\!\! \exp\,i\,\left\{
S_{NEW} \left[\chi^n \right]  + 
S_M \left[\chi^n \right]  + 
\int \Delta^n \left[\chi^m \right] \cdot K_n + 
\int \chi^n \cdot J_n 
\right\} \nonumber
\eeqq 
which means that the quantum average $\langle \Delta^m
\left[x,X;\chi^n \right] \rangle_{J_{_{\chi,K}},K}$ in the presence of the currents $J$ and $K$
\beq \label{59}
\intx\intX \La^{-4}\, \langle \Delta^m
\left[x,X;\chi^n \right] \rangle_{J_{_{\chi,K}},K} 
\cdot J_m (x,X) = 0
\eeq
vanishes.

Next we define the related quantum effective action in the presence of the current $K$ by 
\beq \label{60}
\Ga\left[\chi, K \right] \equiv {\cal W}\left[J_{_{\chi,K}}, K \right]
- \int \chi^n \cdot J_{n\,_{\chi,K}}
\eeq
with the connected vacuum persistence amplitude ${\cal W}\left[J, K \right]$ related to ${\cal Z}\left[J, K \right]$ by
\beq \label{61}
{\cal W}\left[J, K \right] \equiv -i\, \log {\cal Z}\left[J, K \right].
\eeq

Taking the left variational derivative of the effective action w.r.t. a field degree of freedom $\chi^n$ we obtain
\beqq \label{62}
\frac{\de_L \Ga\left[\chi, K \right]}{\de \chi^n (x,X)}
&=& \int \frac{\de_L {\cal W}\left[J, K \right]}{\de J_m (y,Y)}\,
\cdot \frac{\de_L J_{m\,_{\chi,K}} (y,Y)}{\de \chi^n (x,X)} \nonumber \\
&-& J_{n\,_{\chi,K}} (x,X) \\
&-& \int \chi^m (y,Y)\cdot \frac{\de_L J_{m\,_{\chi,K}} (y,Y)}{\de \chi^n (x,X)} \nonumber \\
&=& - J_{n\,_{\chi,K}} (x,X), \nonumber
\eeqq
where we have made use of the definition of $J_{_{\chi,K}}$ which is the current required to give the fields the expectation value $\chi$ in the presence of $K$
\beq \label{63}
\chi^n (x,X) 
= \int \frac{\de_L {\cal W}\left[J, K \right]}{\de J_n (x,X)} 
_{\mid_{_{J = J_{_{\chi,K}}}}}.
\eeq

Taking next the right variational derivative of the effective action w.r.t. the external current $K_n$ we obtain
\beqq \label{64}
\frac{\de_R \Ga\left[\chi, K \right]}{\de K_n (x,X)}
&=& \frac{\de_R {\cal W}\left[J, K \right]}{\de K_n (x,X)}
\nonumber \\
&+& \int \frac{\de_R {\cal W}\left[J, K \right]}{\de J_m (y,Y)}\,
\cdot \frac{\de J_{m\,_{\chi,K}} (y,Y)}{\de K_n (x,X)} \nonumber \\
&-& \int \chi^m (y,Y)\cdot \frac{\de J_{m\,_{\chi,K}} (y,Y)}{\de K_n (x,X)} \\
&=& \frac{\de_R {\cal W}\left[J, K \right]}{\de K_n (x,X)} 
_{\mid_{_{J = J_{_{\chi,K}}}}} \nonumber \\
&=& -i\, \frac{1}{{\cal Z}}\, 
\frac{\de_R {\cal Z}\left[J, K \right]}{\de K_n (x,X)}
_{\mid_{_{J = J_{_{\chi,K}}}}} \nonumber \\
&=& \langle \Delta^n \left[x,X;\chi \right] \rangle_{J_{_{\chi,K}},K}. 
\nonumber
\eeqq 

Finally, inserting both Eqn.(\ref{62}) and Eqn.(\ref{64}) into Eqn.(\ref{59}) we obtain 
\beq \label{65}
\intx\intX \La^{-4}\, 
\frac{\de_R \Ga\left[\chi, K \right]}{\de K_n (x,X)}\cdot 
\frac{\de_L \Ga\left[\chi, K \right]}{\de \chi^n (x,X)}
= 0,
\eeq
i.e. the Zinn-Justin equation for the gauge theory of volume-preserving diffeomorphisms.

Defining the antibracket of two functionals $F\left[\chi, K \right]$ and $G\left[\chi, K \right]$ w.r.t to the field $\chi^n$ and the current $K_n$
\beqq \label{66}
\left( F,\, G \right) &=& \intx\intX \La^{-4}\Bigg\{ 
\frac{\de_R F\left[\chi, K \right]}{\de \chi^n (x,X)}\cdot 
\frac{\de_L G\left[\chi, K \right]}{\de K_n (x,X)} \\
& &\quad\quad -\, \frac{\de_R F\left[\chi, K \right]}{\de K_n (x,X)}\cdot 
\frac{\de_L G\left[\chi, K \right]}{\de \chi^n (x,X)}
\Bigg\} \nonumber
\eeqq
the Zinn-Justin equation finally can be re-written in the form
\beq \label{67}
\left( \Ga,\, \Ga \right) = 0
\eeq
as the interchange of $\chi^n$ and $K_n$ in Eqn.(\ref{65}) simply results in a change of sign.

It is this equation which contains the information at the quantum level related to the original gauge symmetry at the classical level which constrains the form of the ultraviolet divergences so that the theory turns out to be renormalizable - the proof of which we turn to next, adapting the general renormalization proof for Yang-Mills theories as outlined e.g. in \cite{stw2,jez} to our case.

\section{Constraints put on the Perturbative Expansion of the Quantum Effective Action by the Zinn-Justin Equation}

In this section we analyze the constraints on the perturbative expansion of the quantum effective action imposed by the Zinn-Justin equation. They result in a combination of the renormalized action $S_R$ and the infinite part of the $N$-th loop contribution $\Ga_{N,\infty}$ to $\Ga$ being invariant under nilpotent transformations related to, but different from the original BRST-type transformations.

We start by rewriting the action $S [\chi, K]$
\beqq \label{68}
S [\chi, K] &=& S_{NEW} [\chi] + S_M [\chi] + \int  \Delta^n\cdot K_n \\
&=& S_R [\chi, K] + S_\infty [\chi, K] \nonumber
\eeqq
as the sum of a renormalized action $S_R [\chi, K]$ in which masses and coupling constants take their renormalized values and counterterms $S_\infty [\chi, K]$ which cancel the infinities from the loop graphs. Both must be invariant under the same symmetries as the original action $S [\chi, K]$.

Next we turn to the quantum effective action $\Ga [\chi, K]$ which we perturbatively expand in a series
\beq \label{69}
\Ga [\chi, K] = \sum_{N = 0}^\infty \Ga_N [\chi, K]
\eeq
where the $N$-th order contribution $\Ga_N [\chi, K]$ contains both diagrams with $N$ loops as well as diagrams with $N-M$ loops plus counterterms from $S_\infty [\chi, K]$ that cancel infinities in graphs with $M$ loops, where $1 \leq M \leq N - 1$.

To prove the renormalizability of the gauge theory of volume-preserving diffeomorphisms it is sufficient to demonstrate that the infinite parts of the $N$-th order contribution to the quantum effective action display the same terms as the original action $S [\chi, K]$. Hence they can be cancelled by counterterms in $S_\infty [\chi, K]$ of the same form as the ones in the Lagrangian Eqn.(\ref{37}) we have started with.

To do so we start inserting the perturbation series Eqn.(\ref{69}) into the Zinn-Justin equation Eqn.(\ref{67})
\beq \label{70}
\sum_{N = 0}^\infty \sum_{M = 0}^\infty \left(\Ga_N,\, \Ga_M \right)
= \sum_{N = 0}^\infty \sum_{N' = 0}^N \left(\Ga_{N'},\, \Ga_{N - N'} \right)
= 0
\eeq
which yields for the $N$-th order
\beq \label{71}
\sum_{N' = 0}^N \left(\Ga_{N'},\, \Ga_{N - N'} \right) = 0.
\eeq
We note that $\Ga_0 [\chi, K] = S_R [\chi, K]$.

Next we assume that for all $M \leq N-1$ the infinities in all the $M$-loop graphs have been cancelled by counterterms in $S_\infty [\chi, K]$. Infinities in Eqn.(\ref{71}) can then only appear in the $N' = 0$ and $N' = N$ terms which are equal. The infinite part of the condition Eqn.(\ref{71}) forces the infinite contribution $\Ga_{N,\infty}$ to $\Ga_N$ to obey
\beq \label{72}
\left(S_R,\, \Ga_{N,\infty} \right) = 0
\eeq
the consequences of which we evaluate below.

Before doing so we recall that the action $S_{NEW} [\chi] + S_M [\chi] + \int  \Delta^n\cdot K_n$ with $S_{NEW} [\chi]$ from Eqn.(\ref{37}) contains products of fields and their derivatives of dimensionality four or less which guarantees renormalizability in the Dyson-sense, hence counterterms necessary to cancel infinities have dimensionality four or less.

To proceed we next need to establish under which (infinitesimal) symmetry transformations of the action 
\beq \label{73}
\chi^n (x,X) \ar \chi^n (x,X) + \ep F^n [x,X; \chi]
\eeq
the quantum effective action is invariant. To do so we repeat the calculation in section 5 assuming that all: the modified action $S_{NEW}$, the matter action $S_M$, $\int  \Delta^n\cdot K_n$ and the path integral measure in ${\cal Z}\left[J, K \right]$ are invariant under the transformation Eqn.(\ref{73}). 

Repeating the short calculation performed in Eqn.(\ref{58}) yields the general result
\beq \label{74}
\int\, \langle F^n \left[\chi^m \right] \rangle_{J_{_{\chi,K}},K} 
\cdot J_n = \int\, \langle F^n \left[\chi^m \right] \rangle_{J_{_{\chi,K}},K} 
\cdot \frac{\de \Ga [\chi, K]}{\de \chi^n} = 0
\eeq

For symmetry transformations which are linear in the fields
\beq \label{75}
F^n \left[x,X;\chi^m \right] = s^n (x,X) + \int t^n\,_m (x,y; X,Y)\cdot
\chi^m (y,Y)
\eeq 
with $ s^n $ and $ t^n\,_m $ field-independent we further obtain
\beqq \label{76}
& & \langle F^n \left[\chi^m \right] \rangle_{J_{_{\chi,K}},K} 
= s^n + \int t^n\,_m\cdot \langle \chi^m \rangle_{J_{_{\chi,K}},K} \\
& &\quad = s^n + \int t^n\,_m\cdot \chi^m
= F^n \left[\chi^m \right] \nonumber
\eeqq
which means that the quantum effective action $\Ga [\chi, K]$ and hence the infinite parts $\Ga_{N,\infty} [\chi, K]$ in a perturbative expansion are invariant under all the linearly realized symmetry transformations
\beq \label{77}
\de_{_F} \Ga = \int\, F^n \left[\chi^m \right] 
\cdot \frac{\de \Ga [\chi, K]}{\de \chi^n} = 0
\eeq
under which the action $S [\chi, K]$ is invariant. These are: spacetime Poincar\'e transformations ($ x^\mu \ar \La^\mu\,_\nu\, x^\nu + a^\mu $ and related field transformations), antighost translations ($\om^*_\al \ar \om^*_\al + c_\al$), the ghost phase transformations ($\om^\al \ar e^{i\al}\, \om^\al, \om^*_\al \ar e^{-i\al}\, \om^*_\al $) and - related to the gauge symmetry - global inner Poincar\'e transformations ($ X^\al \ar X'^\al $ and related field transformations as in Eqns.(\ref{5}) with ${\cal E}$ $x$-independent) as well as inner scale transformations ($ X^\al \ar \rho\, X^\al $ and $ \La \ar \rho\, \La $). We will come back to the scale transformations when discussing restrictions on the regularization of divergent inner momentum integrals.

There are these linearly realized symmetries together with the restriction Eqn.(\ref{72}) imposed by the Zinn-Justin equation that will be sufficient to determine the general form $\Ga_{N,\infty} [\chi, K]$ of the infinite part of the quantum effective action.

Our first step is to use dimensional analysis and ghost number conservation to determine the $K$-dependence of $\Ga_{N,\infty} [\chi, K]$.

We denote the mass dimension of a field by $[..]$. Then for $[\chi^n] \equiv d_n$ we find $[\Delta^n] = d_n + 1$ and $[ K_n ] = 4 - [\Delta^n] = 3 - d_n$. Hence,
\beqq \label{78}
& & [ A_\mu\,^\al ] = [ \om^\al ] = [ \om^*_\al ] = 1,
\nonumber \\
& & [ K_A ] = [ K_\om ] = [ K_{\om^*} ] = 2, \\
& & [ \psi ] = 3/2,\, [ K_\psi ] = 3/2. \nonumber
\eeqq

As a result the dimension-four quantity $\Ga_{N,\infty} [\chi, K]$ is at most quadratic in the $K_n$ and terms quadratic in $ K_\psi $ can contain only one additional field of dimension one.

Turning to analyze the impact of ghost number conservation we denote the ghost quantum number of a field by $|..|$. Then for $|\chi^n| \equiv g_n$ we find $|\Delta^n| = g_n + 1$ and $|K_n| = -g_n - 1$. Hence,
\beqq \label{79}
& & | A_\mu\,^\al | = | \psi | = 0,\, | K_A | = | K_\psi | = -1, \nonumber \\
& & | \om^\al | = 1,\, | K_\om | = -2 \\
& & | \om^*_\al | = -1,\, | K_{\om^*} | = 0. \nonumber
\eeqq

This shows that no terms in $\Ga_{N,\infty} [\chi, K]$ quadratic in $K_n$ are possible but potentially for $ K_{\om^*} $. For that case we note that the BRST-type transformation $\Delta^{\om^*} = -h$ of $\om^*$ is linear in the fields, and recalling Eqn.(\ref{64}) we find that
\beq \label{80}
\frac{\de_R \Ga [\chi, K]}{\de K_{\om^*}\,^\al }
= \langle \Delta^{\om^*}\,_\al \rangle
= -h_\al
\eeq
is independent of $ K_{\om^*}\,^\al $. So $\Ga [\chi, K]$ itself is linear in $ K_{\om^*}\,^\al $ and the infinite part $\Ga_{N,\infty} [\chi, K]$ does not depend on $ K_{\om^*}\,^\al $ for $N > 0$.

As a result of this first step we hence have fully determined the $K$-dependence of $\Ga_{N,\infty}$ to be
\beq \label{81}
\Ga_{N,\infty} [\chi, K] = \Ga_{N,\infty} [\chi, 0]
+ \int {\cal D}_N^n [x,X;\chi]\cdot K_n (x,X)
\eeq 
introducing the new quantities ${\cal D}_N^n [x,X;\chi]$ the significance of which will become clear in step two. Before proceeding we recall that
\beq \label{82}
S_R [\chi, K] = S_R [\chi]
+ \int \Delta^n [x,X;\chi]\cdot K_n (x,X).
\eeq 

Turning to step two we insert Eqns.(\ref{81}) and (\ref{82}) into the Zinn-Justin relation Eqn.(\ref{72}). We obtain two constraints to zeroth 
\beq \label{83}
\int\, \left\{ \Delta^n [x,X;\chi]\cdot
\frac{\de_L \Ga_{N,\infty} [\chi, 0]}{\de \chi^n (x,X)}
+ {\cal D}_N^n [x,X;\chi]\cdot
\frac{\de_L S_R [\chi]}{\de \chi^n (x,X)} \right\} = 0
\eeq
and to first order in $K$
\beq \label{84}
\int\, \left\{ \Delta^n [x,X;\chi]\cdot
\frac{\de_L {\cal D}_N^m [y,Y;\chi]}{\de \chi^n (x,X)}
+ {\cal D}_N^n [x,X;\chi]\cdot
\frac{\de_L \Delta^m [y,Y;\chi]}{\de \chi^n (x,X)} \right\} = 0.
\eeq 

To further extract the content of the two constraints above we next define
\beq \label{85}
\Ga_N^{(\ep)} [\chi] \equiv S_R [\chi]
+ \ep\, \Ga_{N,\infty} [\chi, 0]
\eeq 
and
\beq \label{86}
\Delta_N^{(\ep) n} (x,X) \equiv \Delta^n (x,X)
+ \ep\, {\cal D}_N^n (x,X)
\eeq
with $\ep$ infinitesimal.

Then Eqn.(\ref{84}) together with the nilpotency of the original BRST-type transformations Eqns.(\ref{57}) tells us that the transformations
\beq \label{87}
\de_{\theta^{(\ep)}} \chi^n (x,X) = \theta\, \Delta_N^{(\ep) n} (x,X)
\eeq
are nilpotent as well:
\beq \label{88}
\de_{\theta^{(\ep)}} \Delta_N^{(\ep) n} (x,X) = 0.
\eeq
And Eqn.(\ref{83}) tells us that $\Ga_N^{(\ep)} [\chi]$ is invariant under the transformations Eqn.(\ref{87}):
\beq \label{89}
\de_{\theta^{(\ep)}} \Ga_N^{(\ep)} [\chi] = 0.
\eeq
These informations will allow us to determine the most general form of both the nilpotent transformations $\Delta_N^{(\ep) n}$ and of $\Ga_N^{(\ep)}$ constraining the infinite parts of the quantum effective action sufficiently to prove renormalizability.

\section{Most General Form of the Quantum BRST-Type Transformations 
$ \Delta_N^{(\ep) n} $}

In this section we determine the most general form of $ \Delta_N^{(\ep) n} $ which will amount to renormalizing the BRST-type transformations Eqns.(\ref{57}) under which $S_{NEW} +S_M$ is invariant.

We start noting that the $ \Delta_N^{(\ep) n} $ must have the same linear transformation properties as the $\Delta^n$ under all: spacetime Poincar\'e transformations, antighost translations, ghost phase transformations, global inner Poincar\'e transformations and inner scale transformations. In addition the $ \Delta_N^{(\ep) n} $ have to be local in the fields and must have the same dimensions as the $\Delta^n$.

In the case of the ghost field $\om^\al$ the mass dimensionality of $ \Delta_N^{(\ep) \om} $ is $[ \Delta_N^{(\ep) \om} ] = 2$ so that the following field combinations could occur in $ \Delta_N^{(\ep) \om} $ in principle: $\om \om$, $\om A$, $A A$, $\om \om^*$, $\om^* \om^*$, $\om^* A$, $\pa \om$, $\pa A$ and $\pa \om^*$. Checking against the additional requirements that $ \Delta_N^{(\ep) \om} $ has to have ghost number $| \Delta_N^{(\ep) \om} | = 2$, has to be invariant under inner Lorentz and inner scale transformations and taking into account that the $\om$ anticommute - forcing the addition of a $\nabla$ - leaves $\nabla \om \om$ as the only non-vanishing possibility. Hence
\beq \label{90}
\om^\al \ar \om^\al 
- \theta\, E_N^{(\ep)\al\be}\,\!_{\ga\de}\cdot \La\, \nabla_\be\, \om^\ga \om^\de,
\eeq
with $E_N^{(\ep)\al\be}\,\!_{\ga\de}$ being a constant tensor in inner space, is the most general form of the transformation Eqn.(\ref{87}) compatible with the above requirements.

In the case of the gauge field $ A_\mu\,^\al $ the mass dimensionality of $ \Delta_N^{(\ep) A} $ is again $[ \Delta_N^{(\ep) A} ] = 2$ so that the same field combinations as in the ghost field case could occur in $ \Delta_N^{(\ep) A} $ in principle. Checking against the additional requirements that $ \Delta_N^{(\ep) A} $ has to have ghost number $| \Delta_N^{(\ep) A} | = 1$, has to be invariant under all: spacetime, inner Lorentz and inner scale transformations leaves $\pa \om$ and $\nabla A \om$ as the only possibilities. Hence
\beq \label{91}
A_\mu\,^\al \ar A_\mu\,^\al + \theta\left( B_N^{(\ep)\al}\,\!_\be\cdot \pa_\mu \om^\be
+ D_N^{(\ep)\al\be}\,\!_{\ga\de}\cdot \La\, \nabla_\be A_\mu\,^\ga \om^\de \right),
\eeq
with $ B_N^{(\ep)\al}\,\!_\be $ and $ D_N^{(\ep)\al\be}\,\!_{\ga\de} $ being constant tensors in inner space, is the most general form of the transformation Eqn.(\ref{87}) compatible with the above requirements.

In the case of the matter field $ \psi $ the mass dimensionality of $ \Delta_N^{(\ep) \psi} $ is $[ \Delta_N^{(\ep) \psi} ] = 5/2$ so that the following field combinations could occur in $ \Delta_N^{(\ep) \psi} $ in principle: $\psi \om$, $\psi A$, $\psi \om^*$ and $\pa \psi$. Checking against the additional requirements that $ \Delta_N^{(\ep) \psi} $ has to have ghost number $| \Delta_N^{(\ep) \psi} | = 1$, has to be invariant under inner Lorentz and inner scale transformations leaves $\nabla \psi \om$ as the only possibility. Hence
\beq \label{92}
\psi \ar \psi - \theta\, C_N^{(\ep)\al}\,\!_\be\cdot \La\, \nabla_\al \psi\, \om^\be,
\eeq
with $ C_N^{(\ep)\al}\,\!_\be $ being a constant tensor in inner space, is the most general form of the transformation Eqn.(\ref{87}) compatible with the above requirements.

For the antighost field $ \om^*_\al $ and the Nakanishi-Lautrup field $ h^\al $ the transformations Eqns.(\ref{87}) remain trivial so that we find the most general form of $ \Delta_N^{(\ep) n} $ to be
\beq \label{93}
\begin{array}{clcl}
\Delta_N^{(\ep) n}\!\!\! &= B_N^{(\ep)\al}\,\!_\be\cdot
\pa_\mu \om^\be + D_N^{(\ep)\al\be}\,\!_{\ga\de}\cdot \La\, \nabla_\be A_\mu\,^\ga \om^\de &\mbox{for}& \, \chi^n = A_\mu\,^\al \nonumber \\
\Delta_N^{(\ep) n}\!\!\! &= - h_\al & \mbox{for} 
& \, \chi^n = \om^*_\al \nonumber \\
\Delta_N^{(\ep) n}\!\!\! &= - E_N^{(\ep)\al\be}\,\!_{\ga\de}\cdot \La\, \nabla_\be\, \om^\ga \om^\de & \mbox{for} & \, \chi^n = \om^\al \\
\Delta_N^{(\ep) n}\!\!\! &= 0 & \mbox{for} & \, \chi^n = h^\al \nonumber \\
\Delta_N^{(\ep) n}\!\!\! &= - C_N^{(\ep)\al}\,\!_\be\cdot \La\, \nabla_\al \psi\, \om^\be & \mbox{for} & \, \chi^n = \psi. \nonumber
\end{array}
\eeq

Next we turn to working out the implication of the nilpotency condition Eqn.(\ref{88}) on  the $ \Delta_N^{(\ep) n} $.

For the ghost field $\om^\al$ we have to take into account
\beq \label{94}
E_N^{(\ep)\al\be}\,\!_{\ga\de}\cdot \nabla_\al \nabla_\be\, \om^\ga \om^\de = 0
\eeq 
which follows from $ \de_{\theta^{(\ep)}} \om^\al$ being divergence-free and conclude that
\beq \label{95}
E_N^{(\ep)\al\be}\,\!_{\ga\de} = -E_N^{(\ep)\be\al}\,\!_{\ga\de}.
\eeq 
To build a constant $ E_N^{(\ep)\al\be}\,\!_{\ga\de} $ we have the tensor $\eta^\al\,_\be$ and the pseudo-tensor $ \ep^{\al\be}\,\!_{\ga\de} $ at hands. As there is parity conservation in inner space we can rule out $ \ep^{\al\be}\,\!_{\ga\de} $ and finally have
\beq \label{96}
E_N^{(\ep)\al\be}\,\!_{\ga\de} \propto \eta^\al\,_\ga\, \eta^\be\,_\de
- \eta^\al\,_\de\, \eta^\be\,_\ga
\eeq
or
\beq \label{97}
E_N^{(\ep)\al\be}\,\!_{\ga\de}\cdot \nabla_\be\, \om^\ga \om^\de \propto
- \om^\be\cdot \nabla_\be\, \om^\al.
\eeq 
As a result the transformation Eqn.(\ref{93}) reduces for the ghost field $\om^\al$ to
\beq \label{98}
\om^\al \ar \om^\al 
- \theta\, {\cal Z}_N^{(\ep)}\, \om^\be\cdot \La\, \nabla_\be\, \om^\al
\eeq
which is easily found to be nilpotent as required.

For the gauge field $ A_\mu\,^\al $ we have to take into account
\beq \label{99}
B_N^{(\ep)\al}\,\!_\be\cdot \nabla_\al \pa_\mu \om^\be 
+ D_N^{(\ep)\al\be}\,\!_{\ga\de}\cdot \nabla_\al \La\, \nabla_\be\, A_\mu\,^\ga \om^\de = 0
\eeq 
which follows from $ \de_{\theta^{(\ep)}} A_\mu\,^\al $ being divergence-free and conclude that
\beq \label{100}
D_N^{(\ep)\al\be}\,\!_{\ga\de} = -D_N^{(\ep)\be\al}\,\!_{\ga\de}
\eeq 
and that
\beq \label{101}
B_N^{(\ep)\al}\,\!_\be \propto \eta^\al\,_\be
\eeq 
as $ \eta^\al\,_\be $ is the only constant tensor of rank $2$ available.
Parity conservation leaves us with
\beq \label{102}
D_N^{(\ep)\al\be}\,\!_{\ga\de} \propto \eta^\al\,_\ga\, \eta^\be\,_\de
- \eta^\al\,_\de\, \eta^\be\,_\ga
\eeq
or
\beq \label{103}
D_N^{(\ep)\al\be}\,\!_{\ga\de}\cdot \nabla_\be\, A_\mu\,^\ga \om^\de \propto
- A_\mu\,^\be\cdot \nabla_\be \om^\al + \om^\be\cdot \nabla_\be A_\mu\,^\al 
\eeq 
so that the transformation Eqn.(\ref{93}) reduces for the gauge field 
$ A_\mu\,^\al $ to
\beq \label{104}
\!\!\!\!\!\! A_\mu\,^\al \ar A_\mu\,^\al + \theta\left( {\cal B}_N^{(\ep)}\, \pa_\mu \om^\al
+ {\cal C}_N^{(\ep)}\, ( A_\mu\,^\be\cdot \La\, \nabla_\be \om^\al
- \om^\be\cdot \La\, \nabla_\be A_\mu\,^\al ) \right).
\eeq
Nilpotency requires
\beqq \label{105}
\de_{\theta^{(\ep)}} \!\!\!\!& &\!\!\!\!\!\!\!\!\!\!
\left( {\cal B}_N^{(\ep)}\, \pa_\mu \om^\al
+ {\cal C}_N^{(\ep)}\, ( A_\mu\,^\be\cdot \La\, \nabla_\be \om^\al
- \om^\be\cdot \La\, \nabla_\be A_\mu\,^\al ) \right)
\nonumber \\
&=& \theta\, \Big\{
- {\cal B}_N^{(\ep)}\, {\cal Z}_N^{(\ep)}\, \pa_\mu \om^\be\cdot \La\, \nabla_\be \om^\al \nonumber \\
&-& {\cal B}_N^{(\ep)}\, {\cal Z}_N^{(\ep)}\, \om^\be\cdot \La\, \nabla_\be \pa_\mu \om^\al \nonumber \\
&+& {\cal B}_N^{(\ep)}\, {\cal C}_N^{(\ep)}\, \pa_\mu \om^\be\cdot \La\, \nabla_\be \om^\al \nonumber \\
&+& {\cal C}_N^{(\ep)\, 2}\, A_\mu\,^\ga\cdot (\La\, \nabla_\ga \om^\be)\cdot (\La\, \nabla_\be \om^\al) \nonumber \\
&-& {\cal C}_N^{(\ep)\, 2}\, \om^\ga\cdot (\La\, \nabla_\ga A_\mu\,^\be)\cdot (\La\, \nabla_\be \om^\al) \nonumber \\
&-& {\cal C}_N^{(\ep)}\, {\cal Z}_N^{(\ep)}\, A_\mu\,^\be\cdot (\La\, \nabla_\be \om^\ga)\cdot (\La\, \nabla_\ga \om^\al) \nonumber \\
&-& {\cal C}_N^{(\ep)}\, {\cal Z}_N^{(\ep)}\, A_\mu\,^\be\cdot \om^\ga\cdot (\La\, \nabla_\be \La\, \nabla_\ga \om^\al) \\
&+& {\cal C}_N^{(\ep)}\, {\cal Z}_N^{(\ep)}\, \om^\ga\cdot (\La\, \nabla_\ga \om^\be)\cdot (\La\, \nabla_\be A_\mu\,^\al) \nonumber \\
&+& {\cal C}_N^{(\ep)}\, {\cal B}_N^{(\ep)}\, \om^\be\cdot \La\, \nabla_\be \pa_\mu \om^\al \nonumber \\
&+& {\cal C}_N^{(\ep)\, 2}\, \om^\be\cdot (\La\, \nabla_\be A_\mu\,^\ga)\cdot (\La\, \nabla_\ga \om^\al) \nonumber \\
&+& {\cal C}_N^{(\ep)\, 2}\, \om^\be\cdot A_\mu\,^\ga\cdot (\La\, \nabla_\be \La\, \nabla_\ga \om^\al) \nonumber \\
&-& {\cal C}_N^{(\ep)\, 2}\, \om^\be\cdot (\La\, \nabla_\be \om^\ga)\cdot (\La\, \nabla_\ga A_\mu\,^\al) \nonumber \\
&-& {\cal C}_N^{(\ep)\, 2}\, \om^\be\cdot \om^\ga\cdot (\La\, \nabla_\be \La\, \nabla_\ga A_\mu\, ^\al)
\Big\} \nonumber \\
&{= \!\!\!\!\!^!}& 0 \nonumber
\eeqq
which is fulfilled if
\beq \label{106}
{\cal C}_N^{(\ep)} = {\cal Z}_N^{(\ep)}.
\eeq 
Setting
\beq \label{107}
{\cal B}_N^{(\ep)} \equiv {\cal Z}_N^{(\ep)}\,{\cal N}_N^{(\ep)}
\eeq 
the transformation Eqn.(\ref{93}) for the gauge field $ A_\mu\,^\al $ takes the final form 
\beq \label{108}
A_\mu\,^\al \ar A_\mu\,^\al + \theta\, {\cal Z}_N^{(\ep)}
\left( {\cal N}_N^{(\ep)}\, \pa_\mu \om^\al
+ A_\mu\,^\be\cdot \La\, \nabla_\be \om^\al
- \om^\be\cdot \La\, \nabla_\be A_\mu\,^\al \right).
\eeq

For the matter field $ \psi $ taking into account
\beq \label{109}
C_N^{(\ep)\al}\,\!_\be \propto \eta^\al\,_\be
\eeq
the transformation Eqn.(\ref{93}) reduces to
\beq \label{110}
\psi \ar \psi - \theta\, {\cal G}_N^{(\ep)}\,  \om^\be\cdot \La\, \nabla_\be \psi
\eeq 
which is found to be nilpotent if
\beq \label{111}
{\cal G}_N^{(\ep)} = {\cal Z}_N^{(\ep)}.
\eeq

As a result we find the most general form of $ \Delta_N^{(\ep) n} $ to be
\beq \label{112}
\begin{array}{clcl}
\Delta_N^{(\ep) n}\!\!\! &= {\cal Z}_N^{(\ep)}
\Big( {\cal N}_N^{(\ep)}\, \pa_\mu \om^\al
&\mbox{for}& \, \chi^n = A_\mu\,^\al \nonumber \\
&\quad\quad\quad + A_\mu\,^\be\cdot \La\, \nabla_\be \om^\al
- \om^\be\cdot \La\, \nabla_\be A_\mu\,^\al \Big) \nonumber \\
\Delta_N^{(\ep) n}\!\!\! &= - h_\al & \mbox{for} 
& \, \chi^n = \om^*_\al \nonumber \\
\Delta_N^{(\ep) n}\!\!\! &= - {\cal Z}_N^{(\ep)}\, \om^\be\cdot \La\, \nabla_\be\, \om^\al & \mbox{for} & \, \chi^n = \om^\al \\
\Delta_N^{(\ep) n}\!\!\! &= 0 & \mbox{for} & \, \chi^n = h^\al \nonumber \\
\Delta_N^{(\ep) n}\!\!\! &= - {\cal Z}_N^{(\ep)}\,  \om^\be\cdot \La\, \nabla_\be \psi & \mbox{for} & \, \chi^n = \psi. \nonumber
\end{array}
\eeq
which amounts to a renormalization of the BRST-type transformations Eqns.(\ref{57}) under which $S_{NEW} + S_M$ is invariant.

\section{Most General Form of $ \Ga_N^{(\ep)} $ and of the Infinite Parts of the Quantum Effective Action}

In this section we show that the most general form of $ \Ga_N^{(\ep)} $ consists of the same terms as the action $S_{NEW}$ we have started with. This demonstrates that potential infinities related to divergent spacetime integrals at any order of perturbation theory can be reabsorbed in a renormalized action $S_R$, hence completing the renormalization proof.

We start noting that $ \Ga_N^{(\ep)} $ can be written
\beq \label{113}
\Ga_N^{(\ep)} [\chi] = \intx\intX \La^{-4}\, {\cal L}_N^{(\ep)} [x,X; \chi]
\eeq
in terms of a Lagrangian $ {\cal L}_N^{(\ep)} $ which is local and in which only combinations of the fields and their derivatives with dimensions less or equal to four can occur.

In addition $ \Ga_N^{(\ep)} $ has to be invariant under all the linear transformations $S_{NEW} = \intx\intX \La^{-4}\, {\cal L}_{NEW}$ is. Recalling that
\beqq \label{114}
{\cal L}_{NEW} &=& - \frac{1}{4}\, F_{\mu\nu}\,^\al \cdot F^{\mu\nu}\,_\al, \nonumber \\
&-& \pa_\mu \om^*_\al \cdot \pa^\mu \om^\al \\
&-& \pa_\mu \om^*_\al \cdot
\left( A^\mu\,_\be\cdot \La\, \nabla^\be \om^\al - \om^\be\cdot \La\, \nabla_\be A^\mu\,^\al \right) \nonumber \\
&+& h_\al \cdot \pa^\mu A_\mu\,^\al 
+ \frac{\xi}{2}\, h_\al \cdot h^\al \nonumber
\eeqq
the action $S_{NEW}$ is found to be invariant under spacetime Poincar\'e transformations, antighost translations, ghost phase transformations, global inner Poincar\'e transformations and inner scale transformations which has to be true for any matter field action $S_M$ as well.

Next we turn to establish the most general form of $ {\cal L}_N^{(\ep)} $ in terms of combinations of fields and their derivatives of dimension four or less, invariant under the above linear transformations and - most importantly - invariant under the modified nilpotent quantum BRST-type transformations Eqn.(\ref{112}).

First, the ghost number conservation requires ghosts $\om$ to be paired with antighosts $\om^*$ and antighost translation invariance forces antighosts to appear with a derivative $\pa \om^*$. As the mass dimension $[ \om \pa \om^* ] = 3$ in ghost terms only one more derivative $\pa$ or gauge field $A$ can appear. Hence, the only two ghost combinations fulfilling all conditions are
\beq \label{115}
\pa_\mu \om^*_\al\, \pa^\mu \om^\be,
\quad
(\pa_\mu \om^*_\al) \La\, \nabla^\be (A^\mu\,_\ga\, \om^\de),
\eeq
where we note that a gauge field $A$ comes together with an inner derivative $\nabla$ to ensure an even number of inner Lorentz indices.

Second, the above symmetry constraints require Nakanishi-Lautrup fields $h$ with mass dimension $[ h ] = 2$ to appear in combination with one of the following terms $h$, $\pa A$, $A A$. Hence the only three $h$-combinations fulfilling all conditions are
\beq \label{116}
h_\al\, h_\be, \quad 
h_\al\, \pa_\mu A^\mu\,_\be, \quad
h_\al\, \La\, \nabla_\be (A^\mu\,_\ga\, A_\mu\,_\de).
\eeq

Adding constant tensors so as to combine inner indices to yield scalar expressions in inner space and noting that $ \eta^{\al\be} $ is the only tensor of second rank allowed by the symmetry requirements we then find the most general expression for $ {\cal L}_N^{(\ep)} $ to be
\beqq \label{117}
{\cal L}_N^{(\ep)} &=& {\cal L}_{\psi A, N}^{(\ep)}
+ \frac{1}{2}\, \xi_N^{(\ep)}\, h_\al \cdot h^\al \nonumber \\
&+& c_N^{(\ep)}\, h_\al \cdot \pa^\mu A_\mu\,^\al \nonumber \\
&+& e_N^{(\ep) \al\be\ga\de}\cdot h_\al\, \La\, \nabla_\be
(A^\mu\,_\ga\, A_\mu\,_\de) \\
&-& Z_{\om, N}^{(\ep)}\, \pa_\mu \om^*_\al \cdot \pa^\mu \om^\al
\nonumber \\
&-& d_N^{(\ep) \al\be\ga\de}\cdot 
(\pa_\mu \om^*_\al) \La\, \nabla^\be 
(A^\mu\,_\ga\, \om^\de), \nonumber
\eeqq
where $ {\cal L}_{\psi A, N}^{(\ep)} $ denotes terms containing only combinations of gauge and matter fields and their derivatives of dimension four or less.

Variation of $ {\cal L}_N^{(\ep)} $ under the modified BRST-type transformations as in Eqns.(\ref{112}) yields
\beqq \label{118}
\de_{\theta^{(\ep)}} {\cal L}_N^{(\ep)} &=&
\de_{\theta^{(\ep)}} {\cal L}_{\psi A, N}^{(\ep)} 
+ \theta\, \bigg\{ (\pa_\mu h^\al)\cdot (\pa^\mu \om_\al)
\Big( -c_N^{(\ep)}\, {\cal Z}_N^{(\ep)}\, {\cal N}_N^{(\ep)} 
+ Z_{\om, N}^{(\ep)} \Big) \nonumber \\
&+& (\pa_\mu h_\al)
\La\, \nabla_\be (A^\mu\,_\ga\, \om_\de)
\cdot \Big( D_N^{(\ep) \al\be\ga\de} c_N^{(\ep)}\, {\cal Z}_N^{(\ep)} + d_N^{(\ep) \al\be\ga\de} \Big) \nonumber \\
&+& (\pa_\mu \om^*_\al)
\cdot \Big( - {\cal Z}_N^{(\ep)}\, Z_{\om, N}^{(\ep)}\, 
\pa^\mu (\om^\be\cdot \La\, \nabla_\be \om^\al) \nonumber \\
& &\quad\quad\quad\quad\quad\quad + \quad\! {\cal Z}_N^{(\ep)}\, 
{\cal N}_N^{(\ep)}\, d_N^{(\ep)\al\be\ga\de}
\cdot \La\, \nabla_\be ((\pa^\mu \om_\ga) \om_\de) \Big) \\
&+& (\pa_\mu \om^*_\al)\cdot 
d_N^{(\ep) \al\be\ga\de}\cdot
{\cal Z}_N^{(\ep)}\, \La\, \nabla_\be \Big(- A^\mu\,_\ga\, \om_\ep\cdot 
\La\, \nabla^\ep \om_\de \nonumber \\
& & \quad\quad\quad\quad\quad\quad 
+ \quad\! (A^\mu\,_\ep\cdot \La\, \nabla^\ep \om_\ga 
- \om^\ep\cdot \La\, \nabla_\ep A^\mu\,_\ga) \om_\de \Big) \nonumber \\
&+& h_\al\cdot e_N^{(\ep) \al\be\ga\de}\cdot
\La\, \nabla_\be \Big( 2 {\cal Z}_N^{(\ep)}\, {\cal N}_N^{(\ep)}\, A^\mu\,_\ga\, \pa_\mu \om_\de \Big) \nonumber \\
&+& h_\al\cdot e_N^{(\ep) \al\be\ga\de}\cdot
\La\, \nabla_\be \Big( 2 {\cal Z}_N^{(\ep)}\, A_{\mu\ga}\, (A^\mu\,_\ep\cdot \La\, \nabla^\ep \om_\de - \om^\ep\cdot \La\, \nabla_\ep A^\mu\,_\de) \Big) \bigg\} \nonumber \\
&{= \!\!\!\!\!^!}& 0. \nonumber
\eeqq 

The first line above vanishes identically if
\beq \label{119}
c_N^{(\ep)} = \frac{Z_{\om, N}^{(\ep)}}{{\cal Z}_N^{(\ep)}\, {\cal N}_N^{(\ep)}},
\eeq
the second line if
\beq \label{120}
d_N^{(\ep) \al\be\ga\de} = -c_N^{(\ep)}\, {\cal Z}_N^{(\ep)}\,
D_N^{(\ep) \al\be\ga\de}
= \frac{Z_{\om, N}^{(\ep)}}{{\cal N}_N^{(\ep)}}\, \Big( \eta^{\al\de}\, \eta^{\be\ga}
- \eta^{\al\ga}\, \eta^{\be\de} \Big)
\eeq
and the last two lines if and only if
\beq \label{121}
e_N^{(\ep) \al\be\ga\de} = 0.
\eeq 

Insertion of Eqn.(\ref{120}) in the third and fourth lines yields
\beq \label{122}
Z_{\om, N}^{(\ep)}\, \pa^\mu (\om^\be\cdot \La\, \nabla_\be \om^\al) + 
{\cal N}_N^{(\ep)}\,
d_N^{(\ep) \al\be\ga\de}\cdot \La\, \nabla_\be ((\pa^\mu \om_\ga) \om_\de) = 0
\eeq
and in the fifth and sixth lines
\beqq \label{123}
& & d_N^{(\ep) \al\be\ga\de}\cdot \La\, \nabla_\be \Big(- A^\mu\,_\ga\, \om_\ep\cdot \La\, \nabla^\ep \om_\de \\
& &\quad\quad\quad\quad\quad + (A^\mu\,_\ep\cdot \La\, \nabla^\ep \om_\ga 
- \om^\ep\cdot \La\, \nabla_\ep A^\mu\,_\ga) \om_\de \Big)= 0 \nonumber
\eeqq 
as required which leaves us with
\beq \label{124}
\de_{\theta^{(\ep)}} {\cal L}_N^{(\ep)}
= \de_{\theta^{(\ep)}} {\cal L}_{\psi A, N}^{(\ep)} 
{= \!\!\!\!\!^!}\,\,\,\, 0.
\eeq 

Next by inspection of Eqn.(\ref{112}) we note that a modified BRST-type transformation acts on the gauge and matter fields exactly in the same way as the original gauge transformation Eqn.(\ref{57}) with local gauge parameter
\beq \label{125}
{\cal E}_N^{(\ep) \al} = {\cal Z}_N^{(\ep)}\, {\cal N}_N^{(\ep)}\, \theta\, \om^\al
\eeq 
and rescaled inner derivatives
\beq \label{126}
\nabla_\al \ar {\tilde\nabla}_\al \equiv \frac{1}{{\cal N}_N^{(\ep)}}\, \nabla_\al.
\eeq
Eqn.(\ref{124}) then simply tells us that $ {\cal L}_{\psi A, N}^{(\ep)} $ has to be gauge invariant when expressed in terms of rescaled inner derivatives $ {\tilde\nabla}_\al $. In addition it contains only combinations of gauge and matter fields and their derivatives of dimension four or less and is invariant under the linear transformations enumerated above. The only gauge field Lagrangian $ {\tilde{\cal L}}_{A, N}^{(\ep)} $ fulfilling these requirements is
\beq \label{127}
{\tilde{\cal L}}_{A, N}^{(\ep)} = - \frac{Z_{A, N}^{(\ep)} }{4}\, 
{\tilde F}_{\mu\nu}\,^\al \cdot {\tilde F}^{\mu\nu}\,_\al,
\eeq
where
\beqq \label{128}
{\tilde F}_{\mu\nu}\,^\al &=& \pa_\mu A_\nu\,^\al - \pa_\nu A_\mu\,^\al \\
&+& A_\mu\,^\be \cdot \La\, {\tilde\nabla}_\be A_\nu\,^\al 
- A_\nu\,^\be \cdot \La\, {\tilde\nabla}_\be A_\mu\,^\al, \nonumber
\eeqq
and the only matter field Lagrangian $ {\tilde{\cal L}}_{M, N}^{(\ep)} $ is
\beq \label{129}
{\tilde{\cal L}}_{M, N}^{(\ep)} =
- Z_{\psi, N}^{(\ep)}\, \bar\psi\, \ga^\mu (\pa_\mu\rvec + A_\mu\,^\al\cdot \La\, {{\tilde\nabla}\rvec}_\al) \psi - \, m_N^{(\ep)}\, \bar\psi\, \psi.
\eeq

As our final result we find the most general $ {\cal L}_N^{(\ep)} $ containing only combination of fields and their derivatives of dimensions four or less, being invariant under all: spacetime Poincar\'e transformations, antighost translations, ghost phase transformations, global inner Poincar\'e transformations and inner scale transformations and most importantly being invariant under the modified BRST-type transformations Eqn.(\ref{112}) to be
\beqq \label{130}
{\cal L}_N^{(\ep)} &=& {\tilde{\cal L}}_{A, N}^{(\ep)}
- Z_{\om, N}^{(\ep)}\, \pa_\mu \om^*_\al \cdot \pa^\mu \om^\al 
\nonumber \\
&-& Z_{\om, N}^{(\ep)}\, \pa_\mu \om^*_\al \cdot
\left( A^\mu\,_\be\cdot \La\, {\tilde\nabla}^\be \om^\al 
- \om^\be\cdot \La\, {\tilde\nabla}_\be A^\mu\,^\al \right) \\
&+& \frac{Z_{\om, N}^{(\ep)}}{{\cal Z}_N^{(\ep)}\, {\cal N}_N^{(\ep)}}\,
h_\al \cdot \pa^\mu A_\mu\,^\al 
+ \frac{\xi_N^{(\ep)}}{2}\, h_\al \cdot h^\al + {\tilde{\cal L}}_{M, N}^{(\ep)}. \nonumber
\eeqq
Apart from the appearance of a number of new constant coefficients this is exactly the original Lagrangian $ {\cal L}_{NEW} + {\cal L}_M $ we have started with. By adjusting the $N$-th order of the unrenormalized constants in the original bare Lagrangian we can absorb $ {\cal L}_N^{(\ep)} $ so that $\Ga_N^{(\ep)} = S_R $ and $ \Ga_{N,\infty} = 0$ which completes the renormalizability proof of the gauge theory of the volume-preserving diffeomorphism group.

\section{Feynman Rules in the Lorentz Gauge}
In this section we derive the Feynman rules for the gauge theory of volume-preserving diffeomorphisms in the Lorentz gauge as a prerequisite to establish viable regularization schemes for the divergent inner momentum integrals occurring in a loop expansion of the quantum effective action.

In order to analyze the structure of inner momentum space integrals at any order of a loop-wise expansion of the quantum effective action we turn to Feynman diagrams. Hence we need to derive the momentum space Feynman rules for the Lagrangian $ {\cal L}_{MOD} $ for gauge and ghost fields in the Lorentz gauge as in Eqn.(\ref{32}) omitting matter contributions for the sake of simplicity because they do not add new features to the present discussion.

As usual we split $ {\cal L}_{MOD} $ into a free Lagrangian $ {\cal L}_{\sl 0} $ and an interaction Lagrangian $ {\cal L}_{INT} $.

The free Lagrangian ${\cal L}_{\sl 0} $ for the gauge and ghost fields 
\beqq \label{131}
{\cal L}_{\sl 0} &=& - \frac{1}{2}\, 
( \pa_\mu A_\nu\,^\al - \pa_\nu A_\mu\,^\al ) \cdot 
\pa^\mu A^\nu \,_\al \nonumber \\
&-& \frac{1}{2\xi} \, \pa_\mu A^\mu\,_\al \cdot \pa^\nu A_\nu\,^\al \\
&-& \pa_\mu \om^*_\al \cdot \pa^\mu \om^\al \nonumber
\eeqq
can by partial integration be easily brought into the usual quadratic form
\beqq \label{132}
{\cal L}_{\sl 0} &=& - \frac{1}{2}\, A_\mu\,^\al \cdot
{\cal D}_{{\sl 0},\xi}^{\mu\nu}\,_{\al\be}\, A_\nu\,^\be \nonumber \\
&-& \om^*_\ga \cdot {\cal D}_{\sl 0}^\ga\,_\de\, \om^\de.
\eeqq
Above we have introduced the non-interacting gauge and ghost field fluctuation operators
\beqq \label{133}
{\cal D}_{{\sl 0},\xi}^{\mu\nu}\,_{\al\be} &\equiv& \left( -\, \eta ^{\mu\nu} \cdot \pa^2  + \left(1 - \frac{1}{\xi} \right)\,
\pa^\mu \, \pa^\nu \right) \eta_{\al\be} \nonumber \\
{\cal D}_{\sl 0}^\ga\,_\de &\equiv& -\, \pa^2\, \eta^\ga\,_\de.
\eeqq
The corresponding free propagators $G^{\sl 0}$ are defined through
\beqq \label{134}
{\cal D}_{{\sl 0},\xi}^{\mu\rho}\,_{\al\ga} \,
G^{{\sl 0},\xi}_{\rho\nu}\,^{\ga\be} (x,y; X,Y) &=& ^T\!\de_\al\,^\be (X-Y) \, \eta^\mu\,_\nu \, \de^4 (x-y) \nonumber \\
{\cal D}_{\sl 0}^\ga\,_\al \, G_{\sl 0}^\al\,_\de (x,y; X,Y) &=& ^T\!\de^\ga\,_\de (X-Y)  \, \de^4 (x-y),
\eeqq
where
\beq \label{135}
^T\!\de_{\al\be}(X-Y) = \int\! \frac{d^4 K}{(2{\pi})^4}\, \La^4\, e^{-iK(X-Y)}
\left( \eta_{\al\be} - \frac{K_\al K_\be}{K^2}\right)
\eeq 
is the scale-invariant delta function transversal in inner space which is compatible with the constraint that both the gauge and the ghost fields are divergence-free in inner space. Note that the transversal delta-function $^T\!\de_{\al\be}(X-Y)$ naturally arises from canonical quantization using Dirac brackets \cite{stw1}.

After a little algebra we find the momentum space gauge and ghost field propagators to be
\beqq \label{136}
G^{{\sl 0},\xi}_{\mu\nu}\,^{\al\be} (p; P) &=&
\frac{1}{p^2 - i\,\ep} \left(\eta_{\mu\nu} - (1-\xi) \frac{p_\mu p_\nu}{p^2} \right) \left( \eta^{\al\be} - \frac{P^\al P^\be}{P^2}\right)
\nonumber \\
G_{\sl 0}^\ga\,_\de (p; P) &=& \frac{1}{p^2 - i\,\ep}\, 
\left( \eta^\ga\,_\de - \frac{P^\ga P_\de}{P^2}\right).
\eeqq
Note that both propagators are transversal in inner space and that they reduce to $\eta^{\al\be}$ when acting on currents which are divergence-free in inner space. As all currents are in fact divergence-free in inner space the inner degrees of freedom do not propagate and we can replace the projection in specific calculations by
\beq \label{137}
\left( \eta^{\al\be} - \frac{P^\al P^\be}{P^2}\right) \ar \eta^{\al\be}.
\eeq
Note that the spacetime parts of the propagators equal the usual Yang-Mills propagators.

Next we calculate the various vertices related to the interaction Lagrangian $ {\cal L}_{INT} $. 

We start with the tri-linear gauge field self-coupling term
\beq \label{138}
-\, \left(\pa_\mu A_\nu\,^\al - \pa_\nu A_\mu\,^\al\right)\, A^\mu\,_\be \cdot \La\, \nabla^\be A^\nu\,_\al
\eeq
corresponding to a vertex with three vector boson lines. If these lines carry incoming spacetime momenta $p_1$, $p_2$, $p_3$, inner momenta $P_1$, $P_2$, $P_3$ and gauge field indices $\mu \al$, $\nu \be$, $\rho \ga$ the contribution of such a vertex to a Feynman graph is
\beqq \label{139}
&-& \!\! 2\, \La\, P_1^\ga\,\eta^{\al\be}\,( p_{2\,\rho} \eta_{\mu\nu} - p_{2\,\mu} \eta_{\nu\rho}) \nonumber \\
&-& \!\! 2\, \La\, P_2^\al\,\eta^{\be\ga}\,( p_{3\,\mu} \eta_{\nu\rho} - p_{3\,\nu} \eta_{\rho\mu}) \\
&-& \!\! 2\, \La\, P_3^\be\,\eta^{\ga\al}\,( p_{1\,\nu} \eta_{\rho\mu} - p_{1\,\rho} \eta_{\mu\nu}) \nonumber 
\eeqq
with
\beq \label{140}
p_1+p_2+p_3=0,\quad\quad P_1+P_2+P_3=0.
\eeq

The quadri-linear gauge field self-coupling term
\beq \label{141}
-\, \frac{1}{2}\, \left(A_\mu\,^\be \cdot \La\, \nabla_\be A_\nu\,^\al - A_\nu\,^\be \cdot \La\, \nabla_\be A_\mu\,^\al\right)\, A^\mu\,_\ga \cdot \La\, \nabla^\ga A^\nu\,_\al
\eeq
corresponds to a vertex with four vector boson lines. If these lines carry incoming spacetime momenta $p_1$, $p_2$, $p_3$, $p_4$, inner momenta $P_1$, $P_2$, $P_3$, $P_4$ and gauge field indices $\mu \al$, $\nu \be$, $\rho \ga$, $\si \de$ the contribution of such a vertex to a Feynman graph is
\beqq \label{142}
&-& \!\! (\La\, P_1^\ga\, \La\, P_2^\de\, \eta^{\al\be} - \La\, P_2^\de\, \La\, P_3^\al\, \eta^{\be\ga}
+ \La\, P_3^\al\, \La\, P_4^\be\, \eta^{\ga\de} - \La\, P_1^\ga\, \La\, P_4^\be\, \eta^{\al\de})
\nonumber \\
& & \quad\quad\quad\quad\quad\quad\quad\quad\quad\quad \cdot( \eta_{\mu\nu}\eta_{\rho\si} - \eta_{\mu\si}\eta_{\nu\rho}) \nonumber \\
&-& \!\! (\La\, P_1^\de\, \La\, P_2^\ga\, \eta^{\al\be} - \La\, P_1^\de\, \La\, P_3^\be\, \eta^{\al\ga}
+ \La\, P_3^\be\, \La\, P_4^\al\, \eta^{\ga\de} - \La\, P_2^\ga\, \La\, P_4^\al\, \eta^{\be\de})
\nonumber \\
& & \quad\quad\quad\quad\quad\quad\quad\quad\quad\quad \cdot( \eta_{\mu\nu}\eta_{\rho\si} - \eta_{\mu\rho}\eta_{\nu\si}) \\
&-& \!\! (\La\, P_1^\be\, \La\, P_3^\de\, \eta^{\al\ga} - \La\, P_1^\be\, \La\, P_4^\ga\, \eta^{\al\de}
+ \La\, P_2^\al\, \La\, P_4^\ga\, \eta^{\be\de} - \La\, P_2^\al\, \La\, P_3^\de\, \eta^{\be\ga})
\nonumber \\
& & \quad\quad\quad\quad\quad\quad\quad\quad\quad\quad \cdot( \eta_{\mu\rho}\eta_{\nu\si} - \eta_{\mu\si}\eta_{\nu\rho}) \nonumber
\eeqq
with
\beq \label{143}
p_1+p_2+p_3+p_4 = 0,\quad\quad P_1+P_2+P_3+P_4 = 0.
\eeq

Finally, the gauge-ghost field coupling term
\beq \label{144}
-\, \pa^\mu \om^*_\ga \left(A_\mu\,^\de\cdot \La\, \nabla_\de \,\om^\ga -\, \om^\de\cdot \La\, \nabla_\de A_\mu\,^\ga\right)
\eeq
corresponds to a vertex with one outgoing and one incoming ghost line as well as one vector boson line. If these lines carry incoming spacetime momenta $p_1$, $p_2$, $p_3$, inner momenta $P_1$, $P_2$, $P_3$ and field indices $\ga$, $\de$, $\mu \al$ the contribution of such a vertex to a Feynman graph becomes
\beq \label{145}
-\, (\La\, P_2^\al\,\eta^{\ga\de} -\, \La\, P_3^\de\,\eta^{\al\be})\, p_{1\,\mu}
\eeq
with
\beq \label{146}
p_1+p_2+p_3=0,\quad\quad P_1+P_2+P_3=0.
\eeq

In summary, the above propagators and vertices allow us to perturbatively evaluate the quantum effective action of the theory in a loop-wise expansion. In addition they are manifestly covariant w.r.t spacetime Poincar\'e transformations and - related to the gauge symmetry - global inner Poincar\'e transformations. Most importantly they are invariant under inner scale transformations ($ P^\al \ar \rho^{-1}\, P^\al $ and $ \La \ar \rho\, \La $). The latter invariance is manifest as each factor of $P$ comes together with a $\La$ and $ \La\, P $ is scale-invariant.

Note that for any Feynman graph the analogon of the sums over Lie algebra structure constants in Yang-Mills theories in the theory under consideration are integrals over inner momentum space variables.

\section{Regularization of the Divergent Inner Momentum Integrals}
In this section we show that $N$-loop inner momentum integrals arising in the expansion of the quantum effective action factorize into sums of $N$ products of one-loop inner momentum integrals. As a key result we find that any regularization procedure for the divergent one-loop inner momentum integrals compatible with the symmetries of the quantum effective action results in a perturbatively well defined quantum field theory. Finally we specify one such regularization procedure.

According to the Feynman rules stated in section 9 vertices contribute simple monomials in the inner momenta to a Feynman graph so that inner momentum space integrals occurring in the $N$-th loop contribution to the perturbative expansion of the quantum effective action factorize into terms of the form
\beqq \label{147}
& & \La^{2k_1 + 4} \int\! \frac{d^{\sl 4}P_1}{(2{\pi})^4}\,
P_1^{\al^1_1} P_1^{\al^1_2}\dots P_1^{\al^1_{2k_1 -1}} P_1^{\al^1_{2k_1}} \cdot
\nonumber \\
& & \quad\quad\quad\quad\quad\quad\quad\quad\quad\quad \vdots \\
& & \cdot \La^{2k_N + 4} \int\! \frac{d^{\sl 4}P_N}{(2{\pi})^4}\,
P_N^{\al^N_1} P_N^{\al^N_2}\dots P_N^{\al^N_{2k_N -1}} P_N^{\al^N_{2k_N}},
\nonumber
\eeqq
where $ \al^j_{2i} $ denote the inner Lorentz indices of the inner momenta $P_j$ appearing in the $j$-th loop integral $ \int\! \frac{d^{\sl 4}P_j}{(2{\pi})^4} $ with $1 \leq j \leq N$ and $0 \leq 2i \leq 2k_j$. Note that only integrals with an even number of $P$ do not vanish.

As these $N$-loop inner momentum integrals factorize into sums of $N$ products of one-loop inner momentum integrals it is sufficient to evaluate integrals of the form
\beqq \label{148}
& & \La^{2k + 4}\, \intP
P^{\al_{_1}} P^{\al_{_2}}\dots P^{\al_{_{2k -1}}} P^{\al_{_{2k}}} \\
& & \sim \mbox{sym} \left\{\eta^{\al_{_1} \al_{_2}}\dots \eta^{\al_{_{2k -1}} \al_{_{2k}}} \right\} \cdot \La^{2k+4} \intP (-P^2)^k \nonumber.
\eeqq
Above we have extracted the Lorentz structure $ \mbox{sym} \left\{\eta^{\al_{_1} \al_{_2}}\dots \eta^{\al_{_{2k -1}} \al_{_{2k}}} \right\} $ from the integral which is a totally symmetric tensor given in terms of the inner Minkowski metric $\eta$.

The remaining scale-invariant integrals are of the form
\beq \label{149}
\Om_k^\La \sim \La^{2k + 4} \intP (-P^2)^k 
\eeq
where the subscript $k$ counts powers of $-P^2$. These divergent $ \Om_k^\La $ correspond to the eigenvalues of the Casimir operators of the gauge algebra in Yang-Mills theories.

Before turning to the regularization of the $ \Om_k^\La $ we note that the $N$-loop contribution Eqn.(\ref{147}) to the perturbative expansion of the quantum effective action can be rewritten in terms of the 
$ \Om_k^\La $ as
\beqq \label{150}
& & \mbox{sym} \left\{\eta^{\al^1_1 \al^1_2}\dots \eta^{\al^1_{2k_1 -1} \al^1_{2k_1}} \right\}\cdot \Om_{k_1}^\La \cdot
\nonumber \\
& & \quad\quad\quad\quad\quad\quad\quad\quad\quad\quad \vdots \\
& & \cdot \mbox{sym} \left\{\eta^{\al^N_1 \al^N_2}\dots \eta^{\al^N_{2k_N -1} \al^N_{2k_N}} \right\}\cdot \Om_{k_N}^\La. \nonumber
\eeqq

Hence, to get a well defined theory it is sufficient to find a regularization procedure for the $ \Om_k^\La $ that is compatible with the symmetries of the quantum effective action, i.e. is invariant under inner Poincar\'e transformations as well as inner scale transformations.

Stated differently each regularization procedure for the $ \Om_k^\La $ that is compatible with the symmetries of the quantum effective action yields a well-defined quantum field theory belonging to the classical gauge theory of volume-preserving diffeomorphisms.

Next we turn to specifing one such regularization procedure compatible with the symmetries of the quantum effective action. This will be done in three steps. In the first step we will slice the inner momentum Minkowski space into light-like, time-like and space-like shells of invariant lengths, in the second we will discard the space-like shells and in the third we will invariantly regularize the remaining integral over light- and time-like shells making use of the existence of an arbitrary point mass and its rest frame.

Slicing the inner Minkowski space in the first step into light-like, time-like and space-like shells of invariant lengths $-P^2 = M^2,\:\: -\infty \leq M^2 \leq \infty$ which are invariant under proper Lorentz transformations of the inner momentum space we can identically rewrite
\beq \label{151}
\Om_k^\La \sim \La^{2k + 4} \int_{-\infty}^{\infty} dM^2\, M^{2k} \intP\, \de\!\left(M^2 + P^2\right).
\eeq
Second, to regularize $ \Om_k^\La $ in a Lorentz-invariant way we cut off the space-like shells with negative lengths $M^2 < 0$ and split $1 = \theta(P^{\sl 0}) + \theta(- P^{\sl 0})$ so that
\beqq \label{152}
\Om_k^\La &\sim& \La^{2k + 4} \int_0^{\infty} dM^2\, M^{2k}
\intP\, \de\!\left(M^2 + P^2\right) \\
& & \quad\quad\quad \cdot\left( \theta(P^{\sl 0}) + \theta(- P^{\sl 0}) \right) \nonumber
\eeqq
which is a Lorentz-invariant procedure. This cutoff of space-like shells arises naturally from the condition of positivity for the Hamiltonian for the gauge and ghost fields as derived in \cite{chw1} which restricts all fields Fourier-transformed over inner space to have support on the set ${\bf V^+}(P)\cup {\bf V^-}(P)$, where
\beq \label{153}
{\bf V^\pm}(P) = \{P\in {\bf M^{\sl 4}}\mid -P^2 \geq 0,\: \pm P^{\sl 0} \geq 0\}
\eeq
denote the forward and backward light cones in inner momentum space.

Third, we can always assume the existence of a point mass which is at rest in some Lorentz frame. In that frame we can define a time-like vector $L^\al$ setting $L^\al = (\La^{-1},\underline 0)$ which has invariant length $L^2 = -\La^{-2}$. Then for all $P \in {\bf V^\pm}(P)$ $ \theta (P^{\sl 0}) $ can be rewritten in an Lorentz invariant way
\beq \label{154}
\theta (P^{\sl 0}) = \theta (\La^{-1}\,P^{\sl 0}) = \theta (-L\cdot P).
\eeq
In addition
\beq \label{155}
-L^2 \pm 2\, L\cdot P = \La^{-2} \mp 2\,\La^{-1}\,P^{\sl 0}
\eeq
again for all $P \in {\bf V^\pm}(P)$.

This allows us to {\it define} $ {\Om\!\!\!\!\!_{^{^{reg}}}}_k^\La $ as a scale-invariant integral over the forward cone ${\bf V^+}(P)$ with a cutoff for $P^{\sl 0} = \sqrt{M^2 + {\underline P}^2} \leq \frac{1}{2\,\La}$ and over the backward cone ${\bf V^-}(P)$ with a cutoff for $P^{\sl 0} = -\sqrt{M^2 + {\underline P}^2} \geq -\frac{1}{2\,\La}$ for fixed $M$ first and then summing over all $M \leq \frac{1}{2\,\La}$
\beqq \label{156}
{\Om\!\!\!\!\!_{^{^{reg}}}}_k^\La &\equiv& \La^{2k + 4} \int_0^{\frac{1}{4\, \La^2}}
\!\!\!\! dM^2\, M^{2k} \intP\, \de\!\left(M^2 + P^2 \right) \\
& & \cdot\left(
\theta(-L\cdot P) \theta(-L^2 + 2\, L\cdot P) +
\theta(L\cdot P) \theta(-L^2 - 2\, L\cdot P)\right) \nonumber \\
&=& \frac{1}{(2\pi)^3\,4^{k + 2}}\, \int_0^1
\!\! dx\, x^k\, \left(\sqrt{1 - x} - x \ln
\frac{\sqrt{1 - x} + 1}{\sqrt x} \right)
\nonumber
\eeqq
which is a positive, finite and manifestly scale-invariant Lorentz scalar for all $k$. Explicitly we find $ {\Om\!\!\!\!\!_{^{^{reg}}}}_{\sl 1}^\La = \frac{1}{720\, (4\pi)^3}$ which corresponds to the eigenvalue of the quadratic Casimir operator in Yang-Mills theories occurring e.g. in the one-loop beta function of the pure gauge theory as calculated for the present case in \cite{chw2}
\beq \label{157}
\be(g) = - \frac{g^3}{(2 \pi)^2}\, \frac{11}{3}\,
{\Om\!\!\!\!\!_{^{^{reg}}}}_{\sl 1}^\La
\eeq
demonstrating its asymptotic freedom as well as in the beta function of the theory including all SM fields
\beq \label{158}
\be(g) = +\, \frac{g^3}{(2 \pi)^2}\, 2\,
{\Om\!\!\!\!\!_{^{^{reg}}}}_{\sl 1}^\La
\eeq
which shows that within the SM the gauge quanta are not confined and hence observable.

This completes the proof that the gauge theory of volume-preserving diffeormorphisms can be consistently quantized and turned into a well-defined quantum field theory. In fact we should rather say into a family of well-defined quantum field theories which differ by the choice of regularization procedure for the inner momentum integrals.

\section{Conclusions}
In this paper we have established that the gauge theory of volume-preser- ving diffeomorphisms of an inner four-dimensional space which arises naturally from the assumption that inertial and gravitational mass need not be the same for virtual quantum states can be consistently quantized and turned into a well-defined quantum field theory - or rather into a family of well-defined quantum field theories which differ by the choice of regularization procedure for the inner momentum integrals.

To get there we have first shown that the gauge-fixed action and the path integral measure occurring in the generating functional for the quantum Green functions of the theory obey a BRST-type symmetry. This has allowed us next to demonstrate that the quantum effective action fulfils a Zinn-Justin-type equation which limits the infinite parts of the quantum effective action to have the same form as the gauge-fixed Lagrangian of the theory proving its spacetime renormalizability. Finally based on the theory's Feynman rules we have shown that the divergent inner space integrals related to the gauge group's infinite volume are regularizable in a way consistent with the symmetries of the theory. In this context it is worth noting that as a byproduct viable quantum gauge field theories are not limited to finite-dimensional compact gauge groups as is commonly assumed.

Finally: what has all of this to do with gravity? To answer this question we have analysed the classical limit $ \hbar \ar 0$ of the gauge theory of volume-preserving diffeomorphisms coupled to a matter field. In that process the inner space collapses, the field dependence on inner coordinates disappears and so does the symmetry under volume-preserving diffeomorphisms of the inner space. On the other hand a new symmetry group emerges: the group of coordinate transformations of four-dimensional spacetime and with it General Relativity coupled to a point particle \cite{chw4}. Hence, as is necessary for the interpretation of the present theory as a quantum theory of gravity GR emerges as its classical limit.

This result implies then that the SM of particle physics can be completed to contain the gravitational interaction at the quantum level as well. 

Practically one starts with the renormalizable action for the SM \cite{stw2,cli,tpc}
\beqq \label{159}
& & S_{SM} = - \intx\, {\cal L}_{Gauge} ( B_n^\nu (x), \pa^\mu B_n^\nu (x)) \nonumber \\
& &\quad  - \intx\, {\cal L}_{Matter} ( \psi_m (x), D_B^\mu (x) \psi_m (x)) \\
& &\quad  - \intx\, {\cal L}_{Higgs} ( \phi (x), D_B^\mu (x) \phi (x) ), \nonumber
\eeqq
where for the sake of clarity we reinsert the arguments $x$ (and $X$ below as well) on which the fields depend. Above $B_n^\nu (x) = B_n^\nu\,_a (x)\, T^a$ denote the $SU(3)\times SU(2)\times U(1)$ gluon and electro-weak gauge fields decomposed in terms of their Lie algebra generators $T^a$, $\psi_m (x)$ denote the quark and lepton fields occurring in three generations and $\phi (x)$ the Higgs field. Finally
\beq \label{160}
D_B^\mu (x) = \pa^\mu + \sum_n B_n^\mu\,_a (x)\, T^a
\eeq
denotes the covariant derivative in a suitable gauge algebra representation.

To get the SM coupled to gravity, in short SM+G, we only have to endow each SM field with inner space coordinates $X^\al$
\beq \label{161}
B_n^\nu (x) \ar B_n^\nu (x,X),\quad \psi_m (x) \ar \psi_m (x,X),\quad
\phi (x) \ar \phi (x,X),
\eeq
introduce the gauge field $A^\mu\,_\al (x,X)$ as in section 2 and the corresponding covariant derivative
\beq \label{162}
D_B^\mu (x) \ar D_{A+B}^\mu (x,X) = \pa^\mu + A^\mu\,_\al (x,X)\cdot \La \nabla^\al + 
\sum_n B_n^\mu\,_a (x,X)\, T^a.
\eeq 
The renormalizable action for the SM+G is then simply
\beqq \label{163}
& & S_{SM+G} = -\frac{1}{4} \intx\intX \La^{-4}\, F_{\mu\nu}\,^\al (x,X) \cdot F^{\mu\nu}\,_\al (x,X) \nonumber \\
& &\quad  - \intx\intX \La^{-4}\, {\cal L}_{Gauge} ( B_n^\nu (x,X), 
D_{A}^\mu (x,X) B_n^\nu (x,X)) \\
& &\quad  - \intx\intX \La^{-4}\, {\cal L}_{Matter} ( \psi_m (x,X), 
D_{A+B}^\mu (x,X) \psi_m (x,X)) \nonumber \\
& &\quad  - \intx\intX \La^{-4}\, {\cal L}_{Higgs} ( \phi (x,X), 
D_{A+B}^\mu (x,X) \phi (x,X) ), \nonumber
\eeqq
where the first term is the gauge field action Eqn.(\ref{22}). All amplitudes or other expressions related to observable quantities calculated within the SM+G obviously have to be evaluated in the physical limit as discussed in \cite{chw3}.

It is reassuring that not only the microscopic strong and electro-weak interactions can be described within a renormalizable quantum gauge field theory framework formulated on a priori flat spacetime. In fact gravity at the quantum level can be described by following exactly the same logic, however, the theory gets more complicated due to its non-compact gauge group having an infinite volume. Yet it is still renormalizable. So Nature seems to allow for a consistent, rupture-free picture based on conservation laws and symmetry considerations at least up to energy scales far beyond experimental reach.

Finally, new physics may derive from Eqn.(\ref{165}), for example in the realm of cosmology and the early universe where new light might be shed on unsolved questions arising e.g. around dark energy \cite{stw4}. For sure the quantum gauge field $A^\mu\,_\al (x,X)$ has left its imprints on the early universe e.g. in a gravitational form of the cosmic background radiation which, however, will not obey the simple Planck distribution its electromagnetic cousin does due to the self-interaction of the gravitational field and its asymptotic freedom nature in the absence of other fields.

\appendix

\section{Notations and Conventions}

Generally, ({\bf M}$^{\sl 4}$,\,$\eta$) denotes the four-dimensional Minkowski space with metric $\eta=\mbox{diag}(-1,1,1,1)$, small letters denote space-time coordinates and parameters and capital letters denote coordinates and parameters in inner space.

Specifically, $x^\la,y^\mu,z^\nu,\dots\,$ denote Cartesian spacetime coordinates. The small Greek indices $\la,\mu,\nu,\dots$ from the middle of the Greek alphabet run over $\sl{0,1,2,3}$. They are raised and lowered with $\eta$, i.e. $x_\mu=\eta_{\mu\nu}\, x^\nu$ etc. and transform covariantly w.r.t. the Lorentz group $SO(\sl{1,3})$. Partial differentiation w.r.t to $x^\mu$ is denoted by $\pa_\mu \equiv \frac{\pa\,\,\,}{\pa x^\mu}$.

$X^\al, Y^\be, Z^\ga,\dots\,$ denote inner Cartesian coordinates we can always choose by partially fixing the gauge to so-called Minkowski gauges \cite{chw1}. The small Greek indices $\al,\be,\ga,\dots$ from the beginning of the Greek alphabet run again over $\sl{0,1,2,3}$. They are raised and lowered with $\eta$, i.e. $x_\al=\eta_{\al\be}\, x^\be$ etc. and transform covariantly w.r.t. the inner Lorentz group $SO(\sl{1,3})$. Partial differentiation w.r.t to $X^\al$ is denoted by $\nabla_\al \equiv \frac{\pa\,\,\,}{\pa X^\al}$. 

The same lower and upper indices are summed unless indicated otherwise.

\end{document}